\documentclass[sigconf]{acmart}
\usepackage{hyperref}
\usepackage{url}
\usepackage{paralist}
\usepackage{caption}
\usepackage{graphicx}
\usepackage{listings}
\usepackage{xcolor}
\usepackage{tabularx}
\usepackage{booktabs} 
\usepackage{multirow}
\usepackage{arydshln}
\usepackage{algorithm}
\usepackage{algpseudocode}
\usepackage{marvosym}
\usepackage{subcaption}
\usepackage{pifont}
\usepackage{wrapfig}
\usepackage[toc,page,header]{appendix}
\usepackage{minitoc}
\usepackage{tocloft}

\newcommand\Ours{WAPITI}

\DeclareMathOperator*{\argmax}{\arg\max}

\AtBeginDocument{%
  }

\setcopyright{acmlicensed}
\copyrightyear{2018}
\acmYear{2018}
\acmDOI{XXXXXXX.XXXXXXX}
\acmConference[Conference acronym 'XX]{Make sure to enter the correct
  conference title from your rights confirmation emai}{June 03--05,
  2018}{Woodstock, NY}
\acmISBN{978-1-4503-XXXX-X/18/06}

\begin{document}

\title[WAPITI:A Watermark for Finetuned Open-Source LLMs]{WAPITI: A Watermark for Finetuned Open-Source LLMs}

\author{Lingjie Chen}
\authornote{Both authors contributed equally to this research.}
\email{ljchen21@m.fudan.edu.cn}
\affiliation{%
  \institution{Fudan University}
  \city{Shanghai}
  \country{China}
}

\author{Ruizhong Qiu}
\authornotemark[1]
\affiliation{%
  \institution{University of Illinois Urbana-Champaign}
  \city{Champaign}
  \state{Illinois}
  \country{USA}}
\email{rq5@illinois.edu}

\author{Siyu Yuan}
\affiliation{%
  \institution{Fudan University}
  \city{Shanghai}
  \country{China}}
\email{syyuan21@m.fudan.edu.cn}

\author{Zhining Liu}
\affiliation{%
  \institution{University of Illinois Urbana-Champaign}
  \city{Champaign}
  \state{Illinois}
  \country{USA}}
\email{liu326@illinois.edu}

\author{Tianxin Wei}
\affiliation{%
  \institution{University of Illinois Urbana-Champaign}
  \city{Champaign}
  \state{Illinois}
  \country{USA}}
\email{twei10@illinois.edu}

\author{Hyunsik Yoo}
\affiliation{%
  \institution{University of Illinois Urbana-Champaign}
  \city{Champaign}
  \state{Illinois}
  \country{USA}}
\email{hy40@illinois.edu}

\author{Zhichen Zeng}
\affiliation{%
  \institution{University of Illinois Urbana-Champaign}
  \city{Champaign}
  \state{Illinois}
  \country{USA}}
\email{zhichenz@illinois.edu}

\author{Deqing Yang}
\authornotemark[2]
\affiliation{%
  \institution{Fudan University}
  \city{Shanghai}
  \country{China}}
\email{yangdeqing@m.fudan.edu.cn}

\author{Hanghang Tong}
\authornote{Corresponding author.}
\affiliation{%
  \institution{University of Illinois Urbana-Champaign}
  \city{Champaign}
  \state{Illinois}
  \country{USA}}
\email{htong@illinois.edu}

\renewcommand{\shortauthors}{L. Chen, R. Qiu, V. B\'eranger, S. Yuan, Z. Liu, T. Wei, H. Yoo, Z. Zeng, D. Yang, H. Tong}

\begin{abstract}
 Watermarking of large language models (LLMs) generation embeds an imperceptible statistical pattern within texts, making it algorithmically detectable. 
Watermarking is a promising method for addressing potential harm and biases from LLMs, as it enables traceability, accountability, and detection of manipulated content, helping to mitigate unintended consequences. 
However, for open-source models, watermarking faces two major challenges: 
\begin{inparaenum}[(i)]
\item incompatibility with fine-tuned models
\item vulnerability to fine-tuning attacks.
\end{inparaenum}
In this work, we propose \textbf{\Ours{}}, a new method that transfers watermarking from base models to fine-tuned models through parameter integration.
To the best of our knowledge, we propose the first watermark for fine-tuned open-source LLMs that preserves their fine-tuned capabilities. 
Furthermore, our approach offers an effective defense against fine-tuning attacks. 
We test our method on various model architectures and watermarking strategies. 
Results demonstrate that our method can successfully inject watermarks and is highly compatible with fine-tuned models. 
Additionally, we offer an in-depth analysis of how 
parameter editing influences the watermark strength and overall capabilities of the resulting models.
\footnote{The model and corresponding code will be released upon publication.}
\end{abstract}



\begin{teaserfigure}
\includegraphics[width=\textwidth]{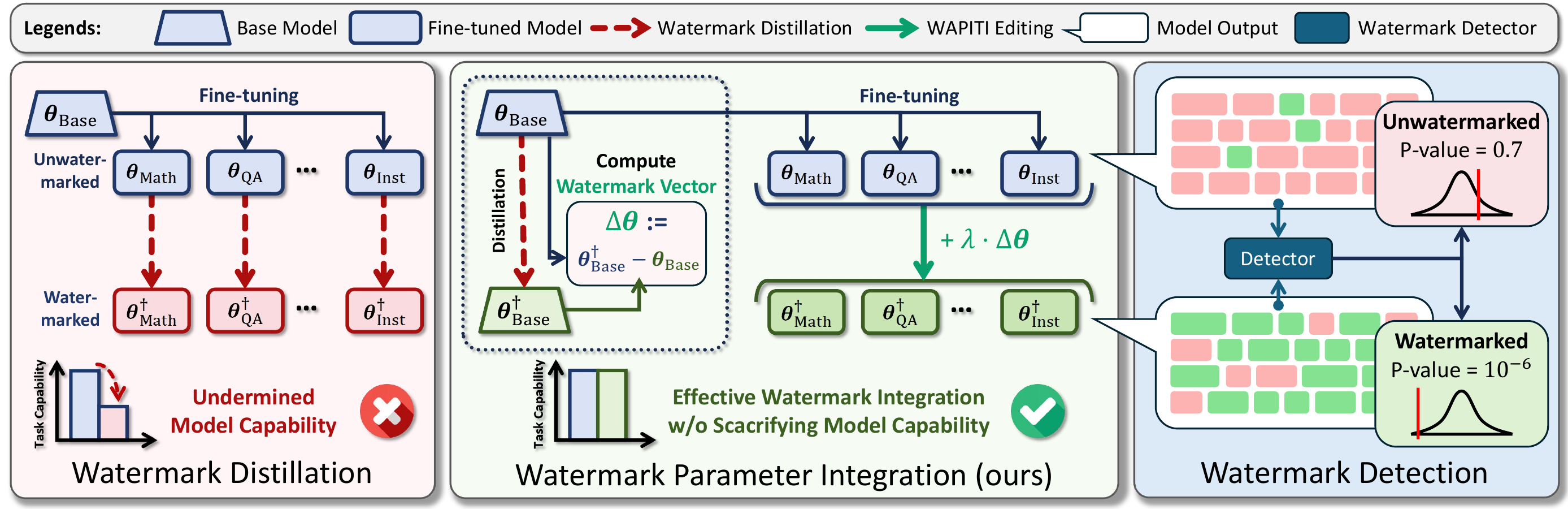} 
\caption{Previous parameter-based watermarking (left) uses distillation which would impair models' fine-tuned capabilities. \Ours{}~(middle) uses watermark-related parameters to transfer watermarking from the base model to fine-tuned models. This method can preserve fine-tuned model capabilities and meanwhile enables them to generate watermarked texts where the green tokens indicate the watermarked tokens (right).}
\label{fig:intro_image}
\end{teaserfigure}

\received{20 February 2007}
\received[revised]{12 March 2009}
\received[accepted]{5 June 2009}

\maketitle

\section{Introduction}
As large language models (LLMs) \cite{touvron2023llama2openfoundation, openai2024gpt4technicalreport} have been integrated into numerous workflows and play an increasingly significant role in everyday life, controlling these LLMs to prevent potential harm has become even more urgent.
Watermarking offers a viable solution by embedding traceable information in model outputs.
It enables the identification of LLM-generated content and can be used to trace back to the source model, serving as a methodological foundation for regulatory oversight of language models.

The vast majority of the prior work on watermarks has focused on closed-source models
~\citep{kirchenbauer2024watermarklargelanguagemodels, aaronson2023watermarking, kuditipudi2024robustdistortionfreewatermarkslanguage}, which are black boxes for users. 
However, with the growing capabilities of open-source models~\citep{touvron2023llama2openfoundation, biderman2023pythiasuiteanalyzinglarge}, the need for oversight of open-source models has become equally important. In other words,
effective watermarking regulation must take both closed-source and open-source models into account to ensure comprehensive oversight and accountability. 

Open-source models release their full parameters to users, and users can fully customize the generation process.
Therefore, users can simply choose an unwatermarked decoding algorithm to evade watermarking, thereby invalidating existing decoding-based watermarking methods.
\cite{gu2024learnabilitywatermarkslanguagemodels} proposed a \emph{parameter-based} method that distills the model using watermarked generations. 
This process, referred to as \emph{watermark distillation}, ensures that the watermarks are retained within the model parameters, preventing users from easily removing them.

However, we observe that this method~\citep{gu2024learnabilitywatermarkslanguagemodels} would impair the fine-tuned capabilities of models, revealing it is not compatible with fine-tuned models.
Additionally, watermark distillation incurs significantly higher computational costs compared to typical fine-tuning.
Furthermore, a severe weakness of parameter-based watermarks is their vulnerability to \emph{fine-tuning attacks}, where malicious users fine-tune the watermarked models with unwatermarked datasets to eliminate their watermarking.
As noted by ~\cite{gu2024learnabilitywatermarkslanguagemodels}, as few as 500 steps of fine-tuning attack can remove the watermark from models.
(See Table~\ref{tab:taxonomy} for overall comparison).

To address these limitations, we propose a new train-free\footnote{"Train-free" means that applying WAPITI to fine-tuned models does not involve any additional model training after acquiring the watermark parameter.} watermarking strategy that transfers watermarks from base models to fine-tuned models (\textbf{\Ours}, \textbf{WA}termark \textbf{P}arameter \textbf{I}n\textbf{T}egrat\textbf{I}on) as shown in Figure~\ref{fig:intro_image}. 
We discover that watermarking bears a similar effect on the output distribution of both base models and fine-tuned models.
The core of our method involves embedding watermarks into models through direct parameter editing, ensuring compatibility with fine-tuned models.
Most importantly, \textbf{\Ours{}} effectively defends against fine-tuning attacks by binding watermarking with the fine-tuning capabilities of the model.


Our main contributions are as follows:
\begin{itemize}
    \item \textbf{Problem.} We identify the incompatibility between current parameter-based watermarking methods and fine-tuned models. 
    Distillation leads to a rapid degradation of fine-tuning capabilities and fails to effectively apply watermarking to models.
    \item \textbf{Method.} To the best of our knowledge, we propose the first watermarking for fine-tuned models (\textbf{\Ours{}}) based on the fact that watermarking causes aligned distribution shift in both base models and fine-tuned models. 
    \item \textbf{Analysis.} 
     We analyze the relationship between watermarking parameters and model performance, revealing how parameter-editing strength affects final outcomes. 
     Furthermore, we establish the relationship between watermarks and the utility of  \Ours{} from a learnability perspective.
    \item \textbf{Evaluation.} \Ours{} achieves high detectability with an AUROC of 0.92 while maintaining near-identical performance on fine-tuning benchmarks for both the Llama-2-7B and Pythia-1.4B families, demonstrating its strong effectiveness and generality.
\end{itemize}

\section{Preliminary}
\subsection{Decoding-based Watermarking}

Large Language Models are generally neural networks based on the transformer architecture, denoted as $f_{\boldsymbol \theta}:\mathcal{V}^* \to \Delta(\mathcal{V})$, which maps a given prefix string $\boldsymbol x \in \mathcal{V}^*$ to a probability distribution over the vocabulary $\Delta(\mathcal{V})$ for predicting the next token, denoted as $f_{\boldsymbol \theta}(\,\cdot \mid \boldsymbol x)$. 
The generation process involves two main steps: \emph{logit generation} followed by \emph{token sampling}~\citep{vaswani2023attentionneed}.

\begin{table*}[t]
\centering
\scriptsize
\begin{tabular}{lcccccc}
\toprule
\multicolumn{1}{l}{\multirow{2}[1]{*}{\textbf{Method}}} & \textbf{Closed-source} & \multicolumn{2}{c}{\multirow{1}[1]{*}{\textbf{Open-sourced}}} & \multicolumn{3}{c}{\multirow{1}[1]{*}{\textbf{Open-sourced Application}}} \\
\cmidrule(lr){2-2}
\cmidrule(lr){3-4}
\cmidrule(lr){5-7}
 & \textbf{LLMs} & \textbf{Base LLMs} &  \textbf{Fine-tuned LLMs} &  \textbf{Efficiency} & \textbf{Vulnerability} & \textbf{Attack Type}  \\
 \midrule
 
 Decoding-based & \ding{51} & \ding{55} & \ding{55} 
 & $\mathcal{O}\left(1\right)$  & Transparent to users &  Directly remove watermark modules\\ %
 Distillation-based & N/A  & \ding{51} & \ding{55}  & $\mathcal{O}\left(\mathcal{C}_{FT}\right)$  & Watermark decay with finetuning & Finetuning attack \\ 
 \textcolor{black}{\textbf{WAPITI}} & N/A & \ding{51} & \ding{51} & \textbf{$\mathcal{O}\left(\mathcal{C}_{FT}/N\right)$} &  \textbf{None} &  \textbf{None} \\ 
 \bottomrule
\end{tabular}
\caption{A taxonomy of LLM watermarking.
"N/A" indicates that the method is not designed for the corresponding setting. 
And $\mathcal{C}_{FT}$ indicates the computation cost of watermark distillation.
$N$ indicates the number of models of the same type in that WAPITI only requires one watermark distillation to watermark all models of the same architecture.
}
\label{tab:taxonomy}
\end{table*}

Decoding-based watermarks are embedded in either stages of generation with the aim of guiding the output distribution toward a targeted direction, incorporating traceable information for detection. 
For instance, KGW~\citep{kirchenbauer2024watermarklargelanguagemodels} increases the frequency of specific tokens during the generation process, and the detector identifies the origin of a text based on the occurrence rate of these tokens.
More specifically, a watermarking algorithm $\mathcal{W}$ employs a watermark key $\phi$ to modify the original next-token distribution $f_{\boldsymbol \theta}(\,\cdot \mid \boldsymbol x)$ into a watermarked version. 
The watermark detector $\mathcal{D}$, using the same watermark key $\phi$, can then retrieve the embedded watermark information.
In general, given a text $x$ and a watermark key $\phi$, the detector $\mathcal{D}$ calculates a p-value for the null hypothesis that the text $x$ is unrelated to $\mathcal{W}$ and $\phi$. 
A text is classified as model-generated if its p-value falls below a predefined threshold.

The key evaluation metrics of watermarking are:
\begin{inparaenum}[(i)]
    \item \textbf{Detectability:} The watermark must ensure that all content generated by the model can be reliably detected by the detector.
    \item \textbf{Utility:} The integration of the watermark should not significantly interfere with the original capabilities of the model.
    \item \textbf{Security:} The watermark should ensure that its hidden pattern within the text is difficult to remove unless a substantial portion of the model output is significantly altered. 
    And for open-source models, the watermark cannot be removed without impairing their capabilities.
\end{inparaenum}


\label{sec:background}
\textbf{Logit-based: KGW} is a watermarking strategy applied directly to output logits of the model (Algorithm 2 in~\cite{kirchenbauer2024watermarklargelanguagemodels}). 
During the next token generation, the vocabulary is pseudorandomly split into green and red lists based on the previous $k$ tokens. 
When $k=0$~\citep{zhao2023provablerobustwatermarkingaigenerated}, the green and red lists are fixed, and when $k\geq1$, the lists are determined by the previous context. 
The green list contains $\gamma\in(0,1)$ proportion of the entire vocabulary, and an additional watermark shift $\delta$ is added to the logits of the tokens in the green list. 
This increases the probability of the green tokens being selected in the final generation. 
During detection, the p-value is calculated by checking whether the proportion of green list tokens exceeds the predefined $\gamma$.


\textbf{Sampling-based: AAR} is the Gumbel softmax scheme from~\cite{aaronson2023watermarking}, which is a special sampling strategy. 
When generating $x_i$, it hashes the previous $k$ tokens using the key $\phi$ to generate a pseudorandom score sequence $\boldsymbol r_i$ for the entire vocabulary $\mathcal{V}$ where $\boldsymbol r_i\in\mathbb R^{|\mathcal V|}$ whose entries are uniformly distributed in $[0,1]$. 
Given the probability distribution $\boldsymbol p_i\in\Delta(\mathcal V)$ of the next token $x_i$, AAR uses Gumbel-Max sampling strategy: $ x_i=\argmax_{j\in \mid\mathcal{V}\mid}(\log p_{i,j}-\log(-\log r_{i,j}))$~\citep{Cane1960IndividualCB},
which introduces some randomness into the sampling stage by adding Gumbel noise $\boldsymbol r_i$.
This sampling strategy would result in watermarked texts having comparative higher score sums. 
During detection, a larger score sum corresponds to a lower p-value against the null hypothesis.
\subsection{Weight-based Watermarking} 
Since the weights of open-source models are fully released, users can modify the decoding method or apply any post-processing to the logits, making decoding-based watermarks easy to remove. 
The most feasible approach\footnote{To the best of our knowledge, this is the only approach for watermarking open-source LLMs that cannot be directly removed by users.} for watermarking is to embed the watermark into the model parameters, enabling LLMs to generate watermarked text under natural sampling distribution. 
Current research~\citep{gu2024learnabilitywatermarkslanguagemodels} has shown that LLMs can learn watermarks via distillation and generate detectable watermarked texts.
By using decoding-based watermark strategies to generate watermarked texts as distillation data,~\cite{gu2024learnabilitywatermarkslanguagemodels} has verified the learnability of multiple watermarks on Llama-2-7B and Pythia-1.4B models. 
However, we found that this parameter-based method is specifically designed for base LLMs. 
In the fine-tuning setting, it significantly impairs the fine-tuned capabilities, as we will demonstrate in \S~\ref{sec:motivation_study}.

\section{Method}
\subsection{Motivating Study}
\label{sec:motivation_study}

\paragraph{Limitation of current weight-based watermarking.}
The current weight-based method enables the base model to generate watermarked texts via distillation.
In this paper, we explore whether the distillation-based approach is compatible with fine-tuned models. Specifically, we ask: can watermark distillation retain the fine-tuned capabilities of the model while embedding the watermark into the fine-tuned model? 
To address this question, we conduct a preliminary experiment.

To obtain a watermarked fine-tuned model using watermark distillation, there are three possible approaches:
\begin{inparaenum}[(i)]
    \item Distilling a fine-tuned model with watermarked content,
    \item Fine-tuning a distilled model that already contains a watermark, or
    \item Fine-tuning a base model using a watermarked fine-tuning dataset.
\end{inparaenum}
We use math-fine-tuned Llama-2-7B and decoding-based watermarking strategies to obtain a watermarked math model.
To counteract the low-entropy nature of mathematical tasks, we use Chain-of-Thought (CoT) reasoning, allowing the model to derive the final answer step by step.
This approach increases the entropy space and enhances the model's ability to detect previous watermarking.
Detailed experimental setups can be found in Appendix~\ref{appendix:motivation_finetuning_setup}.

\begin{figure}[t]
    \centering
    \includegraphics[width=1.0\linewidth]{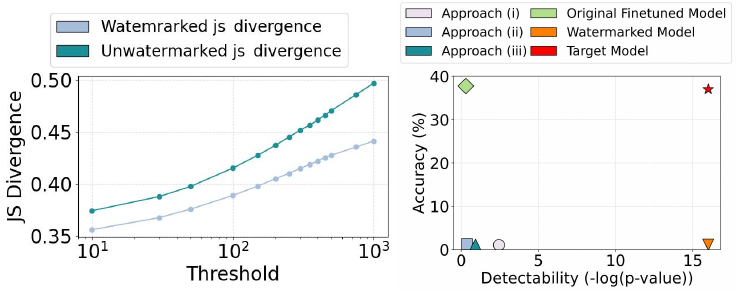}
    \caption{Performance of watermark distillation on fine-tuned models (left) and the distribution alignment of watermarked content compared with unwatermarked content (right) measured with JS divergence.}
    \label{fig:motivation_study}
\end{figure}

Figure~\ref{fig:motivation_study} (Right) compares the watermark detectability (measured by the negative logarithm  of p-value) and fine-tuning utility of the resulting model from all three different approaches.
The utility of the models on GSM8K drops sharply to nearly zero, and the output text shows poor detectability, with a p-value close to the baseline of 0.5.

To better understand this phenomenon, we further analyze the three approaches. 
The first two methods both involve two-phase fine-tuning, which, as studied in previous research, can lead to capability degradation or catastrophic forgetting~\citep{wang2023comprehensivesurveyforgettingdeep}.

For the third method, we believe it holds the most potential to enable the base model to learn mathematical capabilities while embedding the watermark content. 
Therefore, we focus on analyzing the distillation data generated by the math-fine-tuned model and identify two main reasons:
\begin{inparaenum}[(i)] 
    \item The quality of the watermarked math data is inferior to that of the original fine-tuning dataset. 
    Although the generated answers might still be correct, they often contain flawed procedures or random repetitive sequences. 
    Such data can confuse the model and result in a performance decline.
    \item The quantity of watermarked math data is insufficient for the student model to learn the watermark effectively. 
    As noted in ~\cite{gu2024learnabilitywatermarkslanguagemodels}, approximately 320K samples are required for a distilled model to internalize the watermark. 
   With only 7.3k samples in the GSM8K training split and further filtering due to the 40\% accuracy of the model, our final dataset was just 2.2\% of the required size.
    The insufficient distillation data makes the generated outputs lack detectability.
\end{inparaenum}

In a nutshell, our experiments demonstrate that current distillation-based watermarking is incompatible with fine-tuned models. 
This is primarily due to the small size of most fine-tuning datasets, which are insufficient for distillation. 
Additionally, the quality of watermarked samples deteriorates compared to the original ones, leading to a decline in the fine-tuning capabilities of model.

\paragraph{Universal distribution shift from watermarking.}
The primary issue with the current weight-based method is the distillation phase, which underscores the need for a train-free approach to watermark fine-tuned models. 
To this end, we aim to investigate whether there are any similarities between the base models and fine-tuned models when watermarked.


To be specific, we analyze the $n$-gram distribution in the watermarked outputs of both the base and fine-tuned models. 
According to watermarking schemes, $n$-gram could be the smallest meaningful unit, making them a natural starting point. 
Check Appendix~\ref{appendix:ngram_analysis} for detailed justification. 

Our experiment compares the $n$-gram distribution similarities between unwatermarked and watermarked texts generated by the base model and fine-tuned model, respectively. 
We used Llama-2-7B and the math fine-tuned Llama-2-7B~\citep{agarwalla2024enablinghighsparsityfoundationalllama} to generate 640k samples with and without watermarking. 
The watermark used were kgw-k1-gamma0.25-delta2~\citep{kirchenbauer2024watermarklargelanguagemodels} and aar-k2~\citep{aaronson2023watermarking}.

We tokenize all generated text into $n$-grams, where $n$ is determined by the number of tokens used to compute the watermarked next-token probability, as mentioned in \S~\ref{sec:background}.
We then calculate the Jensen--Shannon (JS) 
divergence~\citep{js_divergence} between the watermarked and unwatermarked $n$-grams. 
To reduce noise, we filter out $n$-grams whose frequencies are below a threshold.

The results, shown in Figure~\ref{fig:motivation_study} (Left), indicate that the JS divergence is consistently smaller for watermarked $n$-gram compared to unwatermarked $n$-gram, which suggests that the distribution of watermarked $n$-gram is more similar between base models and fine-tuned models.
This indicates that watermarking distorts the output of both the base and fine-tuned models in similar ways by increasing the frequency of watermarked $n$-grams in the final generation. 

\subsection{Watermark Parameter}
In this section, we focus on deriving the watermarked parameters of fine-tuned models. 
As mentioned in \S\ref{sec:background}, watermarks only perturb the next-token generation $x_t$ according to previous $k$ tokens $x_{t-k},\cdots,x_{t-1}$ and watermark key $\phi$, so that watermark perturbation in next-token probability $f_{\boldsymbol\theta}(\boldsymbol x)$\footnote{For brevity, we identify the next-token probability predictor $f_{\boldsymbol\theta}(\,\cdot\mid \boldsymbol x):\mathcal V\to\mathbb R$ as a vector $f_{\boldsymbol{\theta}}(\boldsymbol x)\in\Delta(\mathcal V)$.} remains the same across different models, where $\boldsymbol x$ is the input prompt.
We denote the watermark perturbation as \( \delta \cdot g(\boldsymbol x) \), where \( \delta \) represents the intensity of the shift, analogous to the watermark shift \( \delta \) in KGW and $g(\boldsymbol x)$ is analogous to the mask of green list in KGW watermarking that indicates which part of vocabulary will be applied watermark shift.
According to experiments in~\ref{appendix:ngram_analysis}, we observe that model parameters can learn watermarking.
Let $\boldsymbol \theta_{\text{Base}}, \boldsymbol \theta_{\text{Base}}^\dagger$ represent parameters of the base model and the watermark-distilled base model respectively.
So we have:  
\begin{align}\label{eqa:watermark_shift_gx}
f_{\boldsymbol{\theta_{\text{Base}}}^\dagger}(\boldsymbol x) = f_{\boldsymbol{\theta_{\text{Base}}}}(\boldsymbol x) + \delta_{\text{Base}} \cdot g(\boldsymbol x).
\end{align}
Similarly, we use \( \boldsymbol{\theta}_{\text{FT}} \) and \( \boldsymbol{\theta}_{\text{FT}}^\dagger \) to represent the parameters of the fine-tuned (FT) models, as well as its watermark-distilled counterpart respectively.
Our ultimate goal is, given an unwatermarked \( \boldsymbol{\theta} \), to find the parameter \( \boldsymbol{\theta}_{\text{FT}}^\dagger \) such that:
\begin{align}\label{eqa:final_goal}
f_{\boldsymbol{\theta}_{\text{FT}}^\dagger}(\boldsymbol{x}) = f_{\boldsymbol{\theta}_{\text{FT}}}(\boldsymbol{x}) + \delta_{\text{FT}} \cdot g(\boldsymbol{x}),
\end{align}
where \( \delta_{\text{FT}} \) is a hyperparameter that controls the watermark detectability.

Let \( \Delta \boldsymbol{\theta_{\text{Base}}}:= \boldsymbol{\theta_{\text{Base}}}^\dagger - \boldsymbol{\theta_{\text{Base}}} \) and \( \Delta \boldsymbol{\theta_{\text{FT}}}:= \boldsymbol{\theta_{\text{FT}}}^\dagger - \boldsymbol{\theta_{\text{FT}}} \) denote the parameter differences introduced by watermark distillation for the base and fine-tuned models, respectively.
We can eliminate \( g(\boldsymbol{x}) \) by substituting \( \boldsymbol{\theta}_{\text{Base}}, \boldsymbol{\theta}_{\text{Base}}^\dagger \) into Eq.~(\ref{eqa:watermark_shift_gx})
and rearranging it as a Taylor expansion. 
$\delta_{\text{Base}}$ denotes the watermark shift of base model:
\begin{align}\label{eqa:watermark_universal_shift}
g(\boldsymbol{x}) &= \frac{1}{\delta_{\text{Base}}} \left( f_{\boldsymbol{\theta}_{\text{Base}}^\dagger}(\boldsymbol{x}) - f_{\boldsymbol{\theta}_{\text{Base}}}(\boldsymbol{x}) \right) \\
&= \frac{1}{\delta_{\text{Base}}} \langle \nabla_{\boldsymbol{\theta}} f_{\boldsymbol{\theta_{\text{Base}}}}(\boldsymbol{x}), \Delta\boldsymbol{\theta}_{\text{Base}} \rangle + O(\|\Delta \boldsymbol{\theta}_{\text{Base}}\|^2).
\end{align}

Furthermore, we observe in Appendix~\ref{appendix:orthogonality_analysis}, that the parameter difference between the fine-tuned model and the base model, $\boldsymbol{\theta_{\text{FT}}} - \boldsymbol{\theta_{\text{Base}}}$, is approximately orthogonal to the parameter difference caused by watermarking, $\boldsymbol{\theta_{\text{Base}}}^\dagger - \boldsymbol{\theta_{\text{Base}}}$: 
\begin{equation}\label{eqa:obs}
   \langle \boldsymbol{\theta}_{\text{FT}} - \boldsymbol{\theta}_{\text{Base}}, \boldsymbol{\theta}_{\text{Base}}^\dagger - \boldsymbol{\theta}_{\text{Base}} \rangle \approx 0.  
\end{equation}
Let $\otimes$ denote the tensor product between differentiation operators, and let $\times_1, \times_2$ denote the mode-1 and mode-2 tensor--matrix product, respectively. Let $\boldsymbol{H}_\text{Base}(\boldsymbol x):=\nabla_{\boldsymbol{\theta}} \otimes \nabla_{\boldsymbol{\theta}} f_{\boldsymbol{\theta}_\text{Base}}(\boldsymbol x)$ be the Hessian. 
As shown in prior studies
, every channel of \( \boldsymbol{H}_{\text{Base}} (\boldsymbol x) \) is approximately the identity matrix \( \boldsymbol{I} \) \citep{jiao2024incontextprobingapproximatesinfluence, yang2024revisitextendenhancehessianfree}. 
Combining it with our observation in Eq.~(\ref{eqa:obs}), we hypothesize that:
\begin{align}\label{eqa:hypo}
     \boldsymbol{H}_{\text{Base}}(\boldsymbol x) \times_1 (\boldsymbol{\theta}_{\text{FT}} - \boldsymbol{\theta}_{\text{Base}})\times_2 (\boldsymbol{\theta}_{\text{Base}}^\dagger - \boldsymbol{\theta}_{\text{Base}}) \approx 0.
\end{align}
The first-order Taylor expansion of \( \nabla_{\boldsymbol{\theta}} f_{\boldsymbol{\theta}_{\text{FT}}} (\boldsymbol x) \) around \( \boldsymbol{\theta} = \boldsymbol{\theta}_{\text{Base}} \) is:
\begin{align}
    \nabla_{\boldsymbol{\theta}} f_{\boldsymbol{\theta}_{\text{FT}}}(\boldsymbol x) 
    &= \nabla_{\boldsymbol{\theta}} f_{\boldsymbol{\theta}_{\text{Base}}}(\boldsymbol x)  \nonumber 
    + \boldsymbol{H}_{\text{Base}}(\boldsymbol x) \times_1 (\boldsymbol{\theta}_{\text{FT}} - \boldsymbol{\theta}_{\text{Base}}) \nonumber \\
    &\quad + O(\|\boldsymbol{\theta}_{\text{FT}} - \boldsymbol{\theta}_{\text{Base}}\|^2)
\end{align}
\begin{align}\label{eqa:taylor_expansion_hessian}
    \boldsymbol{H}_{\text{Base}}(\boldsymbol x) \times_1 (\boldsymbol{\theta}_{\text{FT}} - \boldsymbol{\theta}_{\text{Base}}) \approx \nabla_{\boldsymbol{\theta}} f_{\boldsymbol{\theta}_{\text{FT}}}(\boldsymbol x) - \nabla_{\boldsymbol{\theta}} f_{\boldsymbol{\theta}_{\text{Base}}}(\boldsymbol x).
\end{align}

Next, substituting Eq.~(\ref{eqa:taylor_expansion_hessian}) into Eq.~(\ref{eqa:hypo}), we find that the gradient difference between the fine-tuned and base models, when multiplied by the watermarked parameter difference of base model, is approximately zero:
\begin{align}\label{eqa:inner_produt}
(\nabla_{\boldsymbol{\theta}} f_{\boldsymbol{\theta}_{\text{FT}}}(\boldsymbol x) - \nabla_{\boldsymbol{\theta}} f_{\boldsymbol{\theta}_{\text{Base}}}(\boldsymbol x)) \Delta \boldsymbol{\theta}_{\text{Base}} \approx 0.
\end{align}

By rearranging Eq.~(\ref{eqa:inner_produt}), we conclude that the gradients of the fine-tuned and base models are approximately equal when applied to the watermarked parameter difference:
\begin{align}\label{eqa:gradient_eqa}
\nabla_{\boldsymbol{\theta}} f_{\boldsymbol{\theta}_{\text{FT}}}(\boldsymbol x) \Delta \boldsymbol{\theta}_{\text{Base}} \approx \nabla_{\boldsymbol{\theta}} f_{\boldsymbol{\theta}_{\text{Base}}}(\boldsymbol x) \Delta \boldsymbol{\theta}_{\text{Base}}.
\end{align}
In this way, we obtain the relationship between the gradient of the fine-tuned model and base models. And we now proceed to derive our target $f_{\boldsymbol\theta_{\text{FT}}}(\boldsymbol x)$.
First, by substituting $ g(\boldsymbol{x}) $ from Eq.~(\ref{eqa:watermark_universal_shift}) into Eq.~(\ref{eqa:final_goal}):
\begin{align}\label{eqa:final_goal_2}
f_{\boldsymbol{\theta}_{\text{FT}}^\dagger}(\boldsymbol x) = f_{\boldsymbol{\theta}_{\text{FT}}}(\boldsymbol x) + \left(\frac{\delta_{\text{FT}}}{\delta_{\text{Base}}} \langle \nabla_{\boldsymbol{\theta}} f_{\boldsymbol{\theta}_{\text{Base}}}(\boldsymbol x), \Delta\boldsymbol\theta_{\text{Base}} \rangle + O(\|\Delta\boldsymbol\theta_{\text{Base}}\|^2)  \right).
\end{align}
We define \( \lambda_{\text{FT}} = \frac{\delta_{\text{FT}}}{\delta_{\text{Base}}} \), where \( \delta_{\text{FT}} \) is a hyperparameter, making \( \lambda_{\text{FT}} \) a tunable factor. Next, we substitute the gradient of base model in Eq.~(\ref{eqa:final_goal_2}) with the gradient of fine-tuned model using Eq.~(\ref{eqa:gradient_eqa}):
\begin{align}\label{approximation_final}
f_{\boldsymbol{\theta}_{\text{FT}}^\dagger}(\boldsymbol x) &\approx f_{\boldsymbol{\theta}_{\text{FT}}}(\boldsymbol x) + \langle \nabla_{\boldsymbol{\theta}} f_{\boldsymbol{\theta}_{\text{FT}}}(\boldsymbol x), \lambda_{\text{FT}} \cdot \Delta\boldsymbol\theta_{\text{Base}} \rangle + O\left(\|\Delta \boldsymbol{\theta}_{\text{Base}}\|^2\right), \\
&\approx f_{\boldsymbol{\theta}_{\text{FT}} + \lambda_\text{FT} \cdot \Delta \boldsymbol{\theta}_{\text{Base}}}(\boldsymbol x).
\label{eqa:final_result}
\end{align}
We treat Eq.~(\ref{approximation_final}) as a Taylor expansion of the next-token probability of the model with respect to its parameters.
Based on Eq.~(\ref{eqa:final_result}), we can select:
\begin{align}
\boldsymbol \theta_{\text{FT}}^\dagger := \boldsymbol \theta_{\text{FT}} + \lambda_{\text{FT}} \cdot \Delta  \boldsymbol{\theta}_{\text{Base}}.
\end{align}

\begin{algorithm}[H]
\caption{WAPITI}
\label{alg:method}
\begin{algorithmic}[1]
\renewcommand\algorithmicrequire{\textbf{Input:}}
\renewcommand\algorithmicensure{\textbf{Output:}}
\Require{base model parameter \( \boldsymbol{\theta}_{\text{Base}} \), fine-tuned model parameter \( \boldsymbol{\theta}_{\text{FT}} \), watermark intensity factor \( \lambda_{\text{FT}} \)}
\Ensure{ watermarked fine-tuned model parameter \( \boldsymbol{\theta}_{\text{FT}}^\dagger \)}
    \State \( \boldsymbol{\theta}_{\text{Base}}^\dagger \gets \text{WatermarkDistillation}
    (\boldsymbol{\theta}_{\text{Base}}) \)
    \label{alg:step1}
    \State \( \Delta \boldsymbol{\theta}_{\text{Base}} \gets \boldsymbol{\theta}_{\text{Base}}^\dagger - \boldsymbol{\theta}_{\text{Base}} \)
    \label{alg:step2}
    \State \( \boldsymbol{\theta}_{\text{FT}}^\dagger \gets \boldsymbol{\theta}_{\text{FT}} + \lambda_{\text{FT}} \cdot \Delta \boldsymbol{\theta}_{\text{Base}} \)
    \label{alg:step3}
\end{algorithmic}
\end{algorithm}

According to derivation, we propose \emph{WAtermark Parameter InTegratIon} ({\Ours{}}), which integrates watermark-related parameters of base model to fine-tuned models. The algorithm is shown in Alg.~\ref{alg:method}.
\Ours{} is compatible with various watermarking strategies: after distilling a base model with the desired watermark (Step 1), the watermark can be seamlessly transferred to fine-tuned models without additional costs (Step 3). 
This approach provides an efficient and effective solution for regulating open-source models.



\section{Experiment}

\subsection{Experimental Setup}
\label{sec:experimental_setup}
In this section, we design experiments to evaluate the utility of \Ours{} in two key aspects: \emph{watermark strength} and \emph{fine-tuning ability}, tested across various models and watermarking strategies.

\paragraph{Watermark and hyperparameters.}
We experiment with two representative decoding-based watermarks, KGW and AAR, with different hyperparameters. 
To ensure a fair and consistent comparison, we adopt the same watermarking hyperparameters as used by \cite{gu2024learnabilitywatermarkslanguagemodels}. 
Specifically, for KGW, we set \( k = \{0, 1, 2\} \), \( \gamma = 0.25 \), and \( \delta = \{1, 2\} \); and for AAR, we use \( k = \{2, 3, 4\} \).
The coefficient \( \lambda_{\text{FT}} \) for watermark parameter integration ranges from \( [0, 4] \).

\paragraph{Dataset and model choices.}  
To ensure the generalizability of \Ours{}, we conduct experiments on two widely used LLM families: Llama-2-7B and Pythia-1.4B, which differ in both architecture and parameter Their popularity in the community further ensures that our experiments reflect real-world utility.
We utilize the watermark-distilled base models from~\cite{gu2024learnabilitywatermarkslanguagemodels}.

\paragraph{Evaluation Procedure} 
We evaluate model generation using samples from the RealNewsLike subset of the C4 dataset~\citep{raffel2023exploringlimitstransferlearning}. 
Specifically, the evaluation sample size is 5,000, with a 50-token prompt and a sequence length of 200. We use temperature sampling with $t=1$.
Building on the approach of ~\cite{gu2024learnabilitywatermarkslanguagemodels}, we apply deduplication during post-processing to remove repetitive generations, ensuring the validity of the final detectability results.

\paragraph{Evaluation Metrics} 
To test the compatibility of \Ours{} with fine-tuned models, we focus on three key fine-tuning capabilities: \emph{instruction-following}, \emph{question answering}, and \emph{math}.
We will refer to corresponding fine-tuned models as Llama-chat, Llama-QA, Llama-gsm8k, and Pythia-chat in the experiment results.
Detailed information on the fine-tuned model selection can be found in Appendix~\ref{appendix:finetuned_model_choice}.
The benchmark datasets used are OpenWebText~\citep{Gokaslan2019OpenWeb}, MMLU~\citep{hendrycks2021measuringmassivemultitasklanguage}, and GSM8K~\citep{cobbe2021gsm8kmath}, respectively.

\subsection{Evaluation Metrics}
Following the evaluation methods used in \cite{kirchenbauer2024watermarklargelanguagemodels}, \cite{kuditipudi2024robustdistortionfreewatermarkslanguage}, and \cite{gu2024learnabilitywatermarkslanguagemodels}, we evaluate the models on 5,000 samples drawn from the RealNewsLike subset of the C4 dataset~\citep{raffel2023exploringlimitstransferlearning}. 
The evaluation includes the following metrics:

\paragraph{Watermark detectability.} 
To assess watermark detectability, we compute the median p-value and AUROC (Area Under the Receiver Operating Characteristic Curve), which evaluates the ability to distinguish between watermarked and unwatermarked content. 
The p-value is computed using the z-score method. A lower p-value indicates stronger watermark detectability.
The AUROC is calculated using an equal number of human-generated texts and model-generated watermarked content, both truncated to the same length for consistency.

\paragraph{Generation quality.} 
Generation quality is evaluated using two metrics: perplexity and seq-rep-3 (Sequence Repetition for 3-grams). 
Perplexity provides an overall assessment of the generated text and is calculated using Llama-2-13B. 
Seq-rep-3 measures repetition by calculating the proportion of repeated trigrams~\citep{welleck2019neuraltextgenerationunlikelihood}.


\paragraph{Fine-tuning abilities.} 
To assess whether \Ours{} preserves the fine-tuned capabilities of models, we evaluate the performance of WAPITI fine-tuned models on the following benchmarks:
\begin{inparaenum}[i)]
    \item Question Answering: We use the full MMLU~\citep{hendrycks2021measuringmassivemultitasklanguage} dataset to assess the QA ability of models. This dataset contains approximately 14,000 questions from 57 domains.
    \item Math: We evaluate the model on the test split of GSM8K~\citep{cobbe2021gsm8kmath}, which consists of 1,319 grade-school math word problems designed to assess multi-step reasoning and arithmetic skills.
\end{inparaenum}

\subsection{Results}


\setlength\tabcolsep{5.0pt}
\begin{table*}[t]
\centering
\small
\begin{tabular}{llcccccccc}
\toprule 
    \multicolumn{1}{c}{\multirow{3}[6]{*}{\textbf{Watermark Scheme}}} & \multicolumn{1}{c}{\multirow{3}[7]{*}{{\textbf{Model}}}} & \multicolumn{4}{c}{\footnotesize {Watermark Detectibility}} & \multicolumn{4}{c}{\footnotesize {Generation Quality}} \\ 
  \cmidrule(lr){3-6} \cmidrule(lr){7-10}
     & &  \multicolumn{2}{c}{\multirow{1}[2]{*}{\textbf{p-value($\downarrow$)}}} & 
    \multicolumn{2}{c}{\multirow{1}[2]{*}{\textbf{AUROC($\uparrow$)}}} & 
    \multicolumn{2}{c}{\multirow{1}[2]{*}{\textbf{Perplexity($\downarrow$)}}} & 
    \multicolumn{2}{c}{\multirow{1}[2]{*}{\textbf{seq-rep-3($\downarrow$)}}} \\
    \vspace{-2mm} \\
    \cmidrule(lr){3-4} \cmidrule(lr){5-6} \cmidrule(lr){7-8} \cmidrule(lr){9-10}
     & & \scriptsize{DECO} & \scriptsize{\Ours{}} & \scriptsize{DECO} & \scriptsize{\Ours{}}& \scriptsize{DECO} & \scriptsize{\Ours{}}& \scriptsize{DECO} & \scriptsize{\Ours{}} \\
\cmidrule(lr){1-10}
     \multirow{7}[3]{*}{\textbf{KGW}} 
       & Llama-distilled & $\text{4.2}\!\cdot\!\text{10}^{-\text{25}}$ & \underline{$\text{3.5}\!\cdot\!\text{10}^{-\text{15}}$} & 0.99 & \textbf{0.94} & 5.91 & 5.85  & 0.05 & \underline{0.03} \\
      & Llama-gms8k &  $\text{5.7}\!\cdot\!\text{10}^{-\text{18}}$ & $\text{1.3}\!\cdot\!\text{10}^{-\text{12}}$ & 0.96 & \underline{0.92} & 4.03 & 4.15 & 0.19 & 0.12\\
     & Llama-chat & $\text{1.9}\!\cdot\!\text{10}^{-\text{8}}$  & $\text{7.9}\!\cdot\!\text{10}^{-\text{7}}$ & 0.92 & 0.90 & \textbf{3.12} & \textbf{3.16} & 0.08 & 0.05\\
     & Llama-QA  & $\text{5.1}\!\cdot\!\text{10}^{-\text{13}}$ & $\text{8.1}\!\cdot\!\text{10}^{-\text{7}}$ & 0.96 & 0.91 & \underline{3.50} & 3.44 & 0.08 & 0.04\\
      \cmidrule{2-10}
     & Pythia-distilled & $\text{2.6}\!\cdot\!\text{10}^{-\text{12}}$ & $\text{6.9}\!\cdot\!\text{10}^{-\text{4}}$ & \underline{0.98} & 0.78 & 12.4 & 20.0 & \underline{0.04} & \textbf{0.02}\\
     & Pythia-chat & $\text{5.3}\!\cdot\!\text{10}^{-\text{11}}$ & $\text{1.48}\!\cdot\!\text{10}^{-\text{1}}$ & 0.90 & 0.61 & 7.23 & 6.86 & 0.06 & 0.07\\
  
\midrule
     \multirow{7}[3]{*}{\textbf{AAR}} 
       & Llama-distilled & \underline{$\text{4.2}\!\cdot\!\text{10}^{-\text{88}}$} & $\text{3.6}\!\cdot\!\text{10}^{-\text{12}}$ & \textbf{1.00} & 0.80 & 27.1 & 5.18 & 0.05 & 0.06 \\
      & Llama-gms8k &  \textbf{$\text{6.3}\!\cdot\!\text{10}^{-\text{92}}$} & $\text{6.2}\!\cdot\!\text{10}^{-\text{8}}$ & \textbf{1.00} & 0.77 & 9.13 & 3.73 & 0.15 & 0.14\\
     & Llama-chat & $\text{1.6}\!\cdot\!\text{10}^{-\text{57}}$ & $\text{7.4}\!\cdot\!\text{10}^{-\text{7}}$ & \textbf{1.00} & 0.78 & 20.2 & \underline{3.18} & 0.06 & 0.07\\
     & Llama-QA  & $\text{5.3}\!\cdot\!\text{10}^{-\text{64}}$ & $\text{4.4}\!\cdot\!\text{10}^{-\text{6}}$ & \textbf{1.00} & 0.78 & 5.9 & 3.45 & 0.06 & 0.07\\
      \cmidrule{2-10}
     & Pythia-distilled & $\text{2.0}\!\cdot\!\text{10}^{-\text{73}}$ & \textbf{$\text{7.3}\!\cdot\!\text{10}^{-\text{18}}$} & \textbf{1.00} & 0.85 & 10.5 & 10.8 & \textbf{0.03} & 0.21\\
     & Pythia-chat & $\text{3.3}\!\cdot\!\text{10}^{-\text{66}}$ & $\text{2.08}\!\cdot\!\text{10}^{-\text{1}}$ & \textbf{1.00}  & 0.61 & 10.1 & 9.41 & \textbf{0.03} & 0.07\\
     \midrule
    \multirow{2}[3]{*}{\textbf{None}} 
    & Base Llama & \multicolumn{2}{c}{$\text{4.5}\!\cdot\!\text{10}^{-\text{1}}$} & \multicolumn{2}{c}{0.48} & \multicolumn{2}{c}{3.14} & \multicolumn{2}{c}{0.03} \\
     \cmidrule{2-10}
   & Base Pythia & \multicolumn{2}{c}{$\text{5.6}\!\cdot\!\text{10}^{-\text{1}}$} & \multicolumn{2}{c}{0.49} & \multicolumn{2}{c}{10.3} & \multicolumn{2}{c}{0.04} \\
\bottomrule
\end{tabular}
\caption{Main results for watermark detectability and generation quality of \Ours{} and decoding-based watermarks across different strategies. 
The displayed results represent the average performance, with an integration coefficient of $\lambda_{\text{FT}} = 1$. 
\emph{DECO} refers to the original decoding-based watermark used as the baseline.
}
\label{tab:main_watermarked_results}
\end{table*}



\paragraph{Watermarking results.}
Table~\ref{tab:main_watermarked_results} presents the results of the watermark strength and generation quality of the \Ours{} model.
Since multiple hyperparameter sets were tested for each watermarking strategy, the result table displays the average across all hyperparameter sets for each watermark, with the embedded watermark parameter integration coefficient $\lambda_{\text{FT}}$ fixed to 1.0. 

The results show that \Ours{} effectively transfers the watermark to other models, achieving low p-values and high AUROC scores, indicating strong detectability.
Additionally, the generation quality metrics confirm that \Ours{} preserves the models' original capabilities.
However, the detectability in WAPITI fine-tuned models is slightly lower compared to the watermark-distilled base models, suggesting that some watermarking information is lost during the transfer process.

Of the two watermarks tested, KGW consistently outperforms AAR in watermark transfer, exhibiting higher AUROC scores. 
This trend is also observed in the watermark-distilled models from~\cite{gu2024learnabilitywatermarkslanguagemodels}, which we partly attribute to the complexity of the AAR scheme, as it combines logits with pseudorandom scores. 
A more detailed analysis of this difference is provided in Section~\ref{appendix:coeff_analysis}.
Comparing the performance across different models, the watermark detectability in Pythia models is lower than in Llama models. 
Analyzing the generations of Pythia models suggests that this difference is largely due to the models' inherent capabilities.
Nevertheless, the parameter integration yields p-values significantly below the baseline of 0.5, indicating that watermarking-related knowledge is still injected to a certain degree.
\begin{figure}[t]
    \centering
    \includegraphics[width=1.0\linewidth]{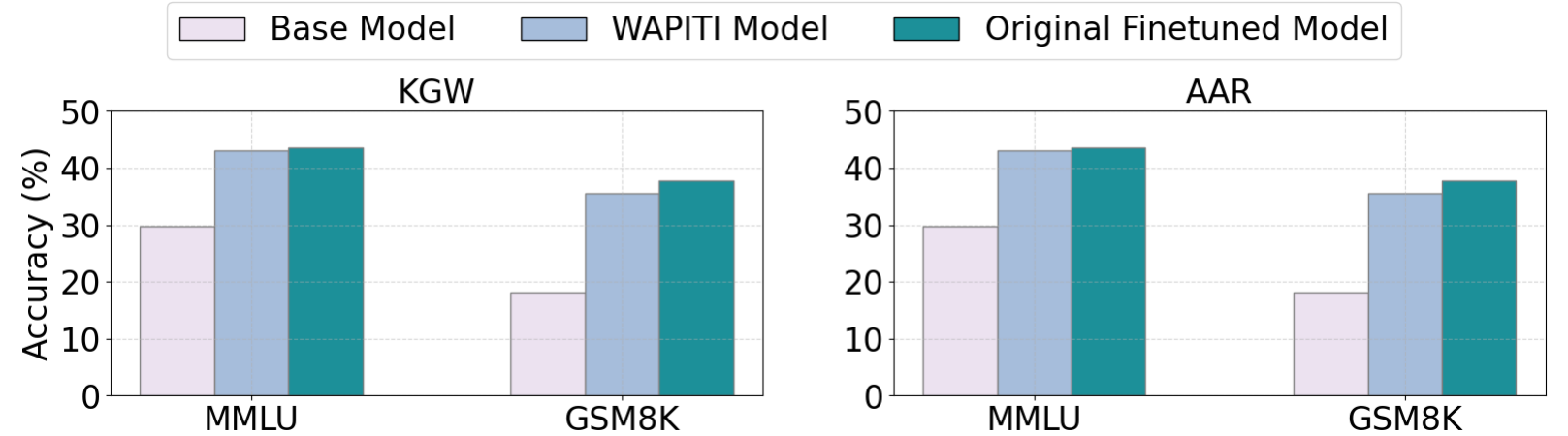}
    \caption{Performance of \Ours{} models on fine-tuning ability benchmarks are intact after watermarking.}
    \label{fig:finetuned_ability_plot}
\end{figure}

\paragraph{Fine-tuned ability results.}
Figure~\ref{fig:finetuned_ability_plot} compares the fine-tuning performance of \Ours{} models with the base model and original fine-tuned models. 
For both QA and Math tasks, \Ours{} models show performance nearly identical to the original fine-tuned models for both KGW and AAR watermarking, demonstrating that \Ours{} effectively preserves the models' original capabilities and is fully compatible with fine-tuned models.

Combined with the results from Table~\ref{tab:main_watermarked_results}, we conclude that \Ours{} is an effective and efficient watermarking method for fine-tuned models, allowing them to retain the watermark while preserving both generation quality and fine-tuned capabilities.

\paragraph{Robustness to Attacks}
WAPITI embeds watermarks through parameter integration, so the watermark parameters must be kept secret, like secret key. 
Otherwise, malicious users could directly remove the integrated parameters to invalidate the watermark.

Beyond this direct attack, we evaluate \Ours{}'s robustness against the classical watermark elimination method: text edits.
Detailed experimental setups and robustness analyses are provided in Appendix~\ref{appendix:robustness_analysis}.


\subsection{Analysis}
In this section, we conduct additional experiments to address three core questions:
\begin{inparaenum}[(i)]
\item How do watermark parameters affect the watermark detectability and capabilities of \Ours{} models?
\item Does \Ours{} offer defense against fine-tuning attacks?
\item Why is \Ours{} compatible with fine-tuned LLMs?
\end{inparaenum}
These experiments provide insights into \Ours{} for improved utilization.

\subsection{How will the watermark parameter affect the model performance?}
\label{appendix:coeff_analysis}


\begin{figure*}[htbp]
\begin{center}
    \includegraphics[width=0.7\textwidth]{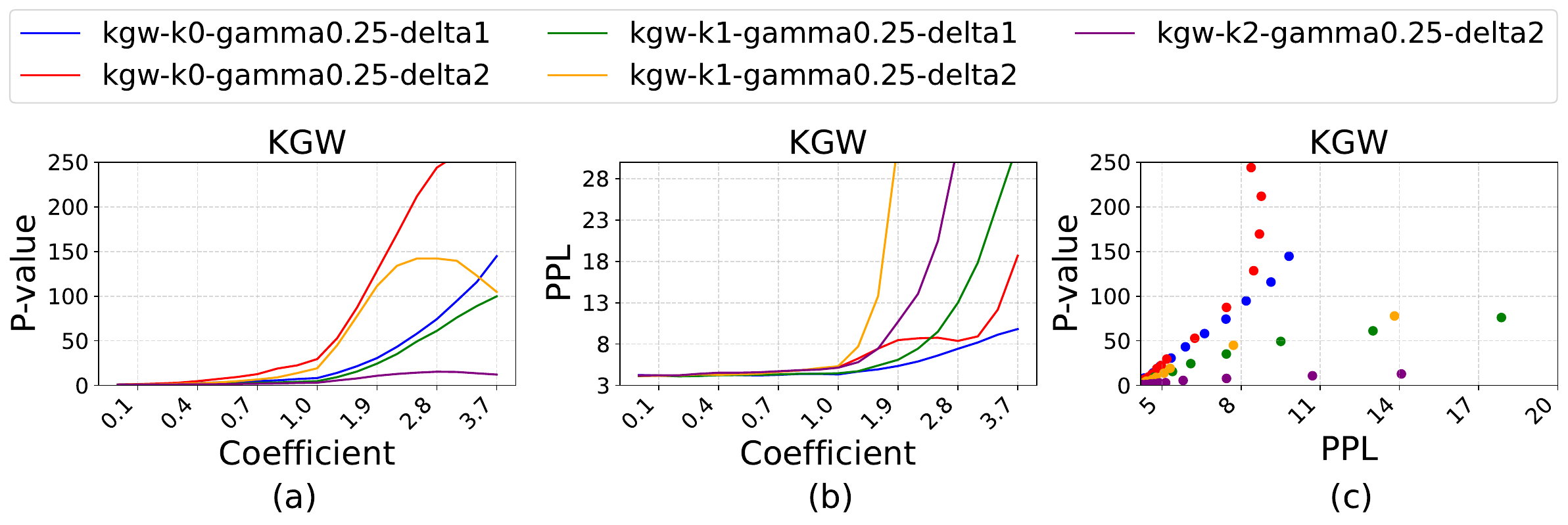}
    \vspace{-4mm} 
    \includegraphics[width=0.7\textwidth]{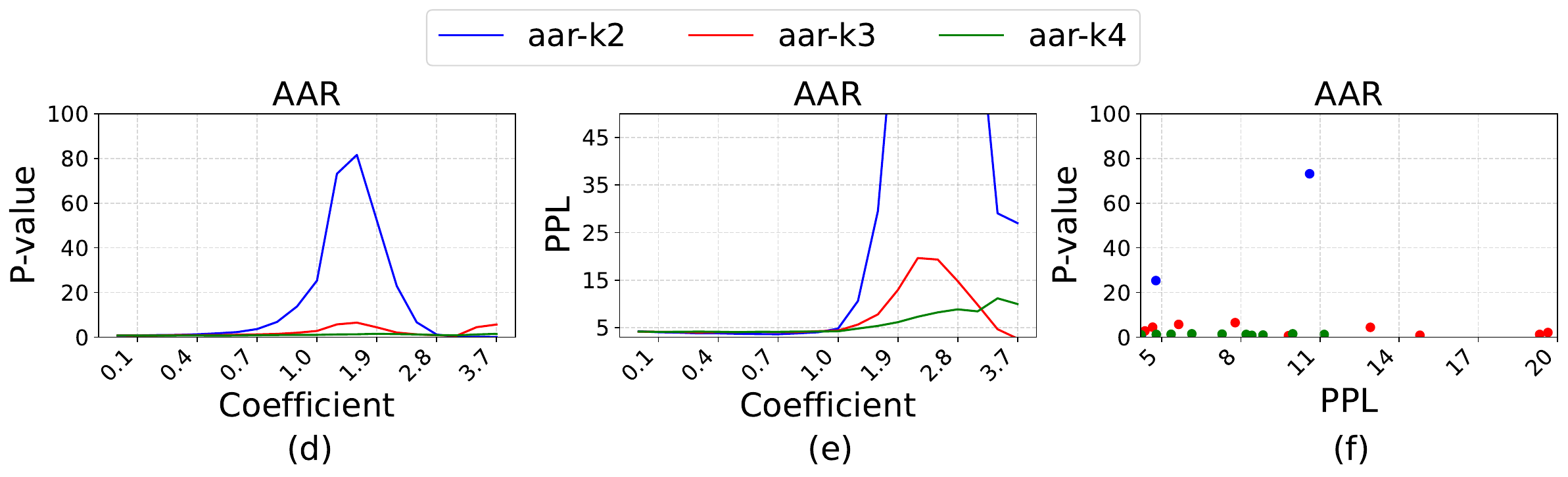}
\end{center}
\caption{Watermark detectability and output perplexity of the \Ours{} model as a function of the watermark integration coefficient $\lambda_{\text{FT}}$ (left and middle). The scatter plot shows the relationship between perplexity and detectability (right).}
\label{fig:coeff_analysis}
\end{figure*}

We evaluated the watermark detectability of the model and generation quality across varying coefficients \( \lambda_{\text{FT}} \) for watermark parameter integration.
Figure~\ref{fig:coeff_analysis} illustrates the watermark detectability (measured by p-value) and perplexity of the \Ours{} Llama-math model at different values of \( \lambda_{FT} \).

The results in Figure~\ref{fig:coeff_analysis}(b)(e) show that when the coefficient is in the range [0, 1], watermark strength increases steadily, while perplexity stays below 5.0, indicating minimal impact on the model's generation capability. 
Additionally, the gradient of watermark strength in Figures~\ref{fig:coeff_analysis}(a) and (d) varies with watermarking hyperparameters. 
For KGW, smaller $k$ reduces the influence of preceding tokens, and larger $\delta$ enhances detectability. 
Similarly, for AAR, a smaller $k$ also reduces the impact of the previous context.
Thus, smaller $k$ and larger $\delta$ make the watermark easier to learn, consistent with~\cite{gu2024learnabilitywatermarkslanguagemodels}. 
These trends are evident in Figure~\ref{fig:coeff_analysis}(a) and (d), where the gradient of watermark strength aligns with watermark learnability.

However, when the coefficient exceeds 1.0, distinct patterns emerge for the two watermarks. 
For KGW, both watermark detectability and perplexity increase monotonically with the coefficient. 
AAR shows a parabolic behavior in both strength and perplexity, with their extrema occurring at different values. 
This divergence suggests that the watermark parameter, while generally applicable, can interfere with the model when $\lambda$ becomes large.


Figure~\ref{fig:coeff_analysis}(c),(f) show scatter plots of perplexity vs. p-value, illustrating the trade-off between watermark detectability and output quality. 
Perplexity is constrained within the [0, 20] range to preserve generation quality. 
KGW exhibits a linear relationship between watermark strength and perplexity, reflecting the expected trade-off. 
In contrast, the AAR plot shows a more chaotic pattern, with no clear correlation. 
This difference stems from the watermarking mechanisms: KGW is explicitly decomposable into n-grams, whereas AAR involves both logits and pseudorandom scores, making it more difficult to learn. 
These results offer insights into the most suitable watermarking strategy for \Ours{}.

\subsection{Can the watermark vector protect fine-tuned abilities?}
\label{appendix:defense_finetune_attack}
\begin{figure}[t]
    \centering
    \includegraphics[width=0.8\linewidth]{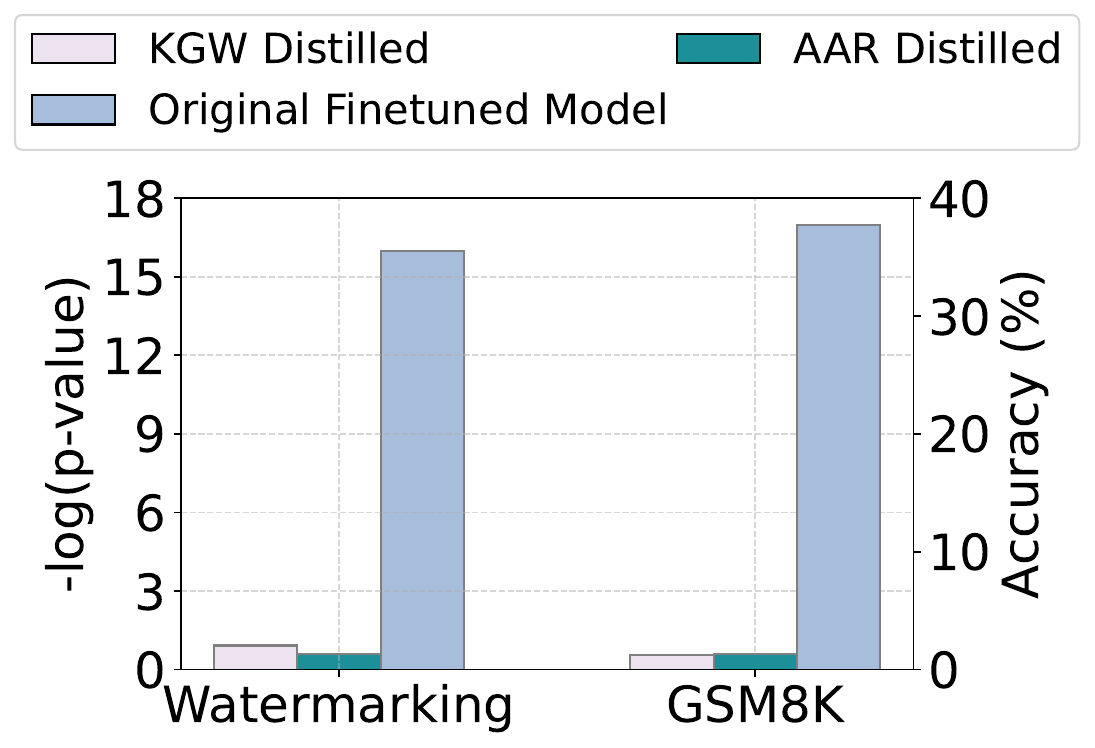}
    \caption{The results reveal that fine-tuning attacks degrade both model performance and watermark removal ability, making them an effective defense.}
    \label{fig:finetuning_attack_defense}
\end{figure}
\paragraph{Fine-tuning Attack}
A critical challenge for weight-based watermarking is defending against fine-tuning attacks. 
Watermark fragility in the face of fine-tuning is particularly difficult to address, as fine-tuning can be viewed as a form of "reverse watermarking." 
Just as distillation can embed a watermark into the model, fine-tuning can potentially remove it, restoring the output distribution of the model to its original state.

Recall the definition in the \S~\ref{sec:background}, the \textbf{Utility} of a watermark is defined by the difficulty of removing it without significantly altering the generated content or impairing the inherent capabilities of the model. 

\paragraph{Fine-tuning Attack Setup} We use fine-tuned models embedded with each watermark type: KGW with $k=0$, $\gamma=2$, and $\delta=2$. These models are then further fine-tuned on the OpenWebText dataset~\citep{Gokaslan2019OpenWeb} for 1,000 steps.
The training configurations are remained the same as~\cite{gu2024learnabilitywatermarkslanguagemodels}.
For each model checkpoint every 200 steps, we generate 200-token completions using prompts of 50 tokens from the C4 RealNewsLike dataset.
Then we calculate the model's fine-tuned capability and its watermark detectability after fine-tuning to show the impact of the fine-tuning attack's impact.

\paragraph{Defense for Fine-tuning Attack} To defend against fine-tuning attacks, we can bind watermarking to the fine-tuned abilities of the model.
In doing so, if malicious users attempt to fine-tune the model to remove the watermark, the fine-tuned capabilities of the model will also be severely compromised, thereby enhancing the robustness of the watermark in fine-tuned models.
In this experiment, we select the Llama-Math and Llama-QA, each embedded with the k0-gamma0.25-delta2 watermark.
Then we test how watermark strength and fine-tuning performance are affected after additional fine-tuning.

As shown in Figure~\ref{fig:finetuning_attack_defense}, both models' watermark detectability and fine-tuning capabilities declined significantly after just 400 steps of fine-tuning attack.
This result demonstrates that \Ours{} on fine-tuned models is robust to fine-tuning attacks, addressing the vulnerabilities of previous watermark distillation methods.

\subsection{How can the watermark vector be compatible with fine-tuned abilities?} 
\label{appendix:orthogonality_analysis}
\begin{figure}[t]
    \centering
    \includegraphics[width=0.7\linewidth]{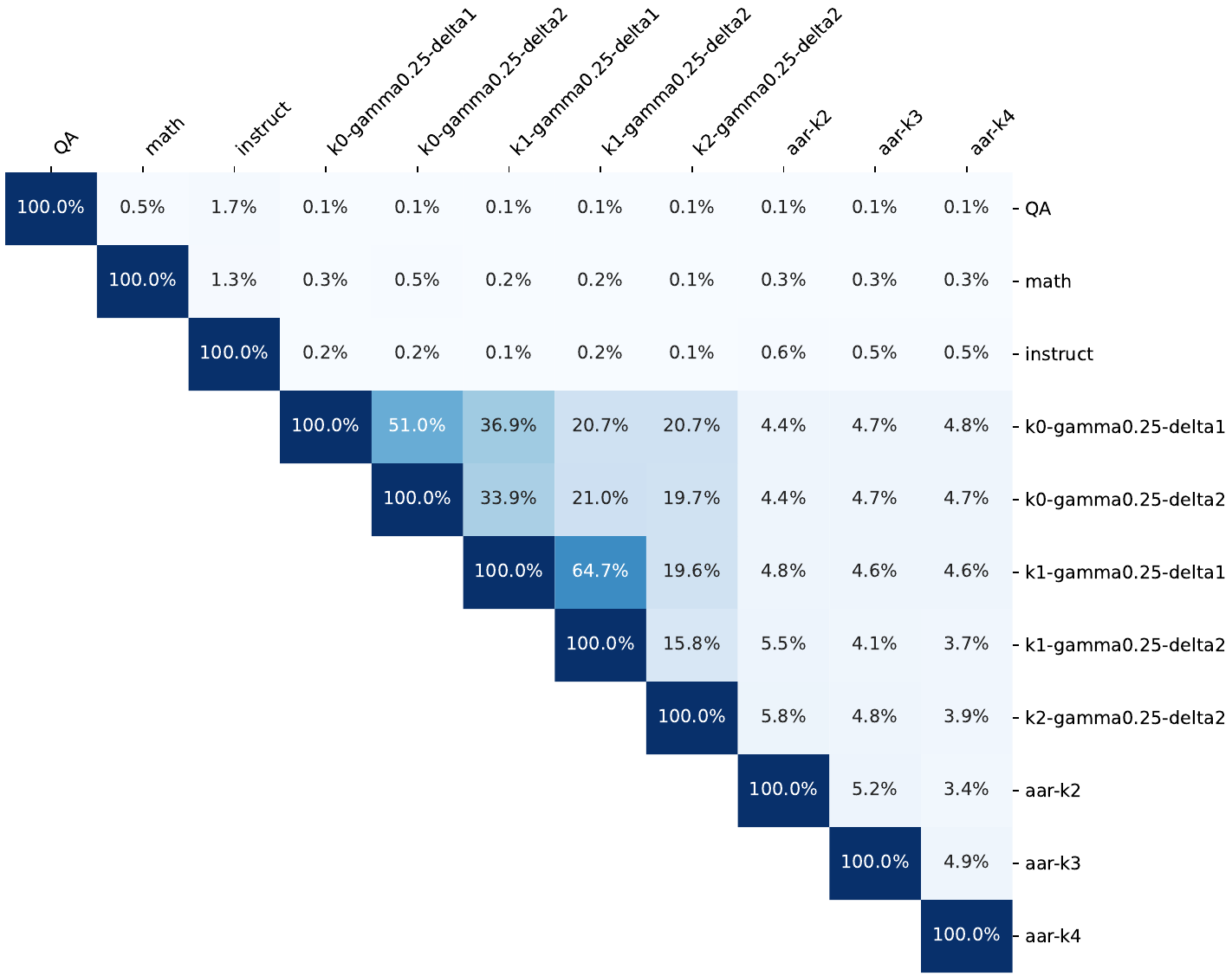}
    \caption{The plot shows cosine similarity, indicating clear orthogonality between watermark parameter differences and fine-tuning parameter differences.}
    \label{fig:watermark_parameter_similarity}
    \vspace{-15pt}
\end{figure}
Finally, We investigate why the \Ours{} is effective across different fine-tuned models by employing a parameter-based approach similar to~\cite{ilharco2023editingmodelstaskarithmetic}.
We calculate the cosine similarity between the watermark parameter and task vectors~\citep{ilharco2023editingmodelstaskarithmetic}, where the task vectors represent the parameter differences between the fine-tuned and base models.

As shown in Figure~\ref{fig:finetuned_ability_plot}, the watermark parameters exhibit strong orthogonality with the fine-tuned parameters, minimizing interference between watermarking and fine-tuning. This likely explains why \Ours{} preserves fine-tuning capabilities.

Additionally, the watermark parameters from different schemes also demonstrate clear orthogonality.
Interestingly, higher similarity is observed within the KGW family, particularly when $k$ values are the same.  
Since identical random seeds and sampling mechanisms are used when $k$ values are the same, this generates identical green lists, leading models to learn the same $n$-grams.
This similarity further indicates that watermark parameters encode specific knowledge about the watermarking schemes.
Overall, this experiment provides strong analytical evidence supporting the effectiveness of \Ours{}.

\section{Related Work}

\paragraph{Text steganography.}
Steganography involves embedding information within texts for the purposes of detection or secret communication.
Steganography methods can be categorized into \emph{edit-based} and \emph{generative} approaches. 
Edit-based methods include rule-based transformations~\citep{DBLP:conf/mediaforensics/WilsonBK14, wilson2016steganograph}, synonym-based substitution~\citep{shirali2008synonymsteganography}, and neural network-based transformations~\citep{fang-etal-2017-generating, abdelnabi2021adversarialwatermarkingtransformertracing, ueoka2021frustratinglyeasyeditbasedlinguistic}. 
On the other hand, generative methods embed information directly during the text generation process~\citep{ziegler2019neurallinguisticsteganography, dai-cai-2019-towards}.

\paragraph{Text watermarking.}
Earlier works in text watermarking typically embedded information through post-processing of texts, closely resembling steganography~\citep{venugopal-etal-2011-watermarking,yang2021tracingtextprovenancecontextaware}.
More recent studies have shifted towards decoding-based watermarking, hiding information by perturbing the text during the decoding phase~\citep{kirchenbauer2024reliabilitywatermarkslargelanguage,aaronson2023watermarking,zhu2024duwakdualwatermarkslarge, krishnaparaphraseattack2023, kuditipudi2024robustdistortionfreewatermarkslanguage,zhao2023provablerobustwatermarkingaigenerated,christ2023undetectablewatermarkslanguagemodels, wu2024resilientaccessibledistributionpreservingwatermark, liu2024adaptivetextwatermarklarge, giboulot2024watermaxbreakingllmwatermark, lu2024entropybasedtextwatermarkingdetection, ren2024robustsemanticsbasedwatermarklarge, wang2024codablewatermarkinginjectingmultibits}. 
Different watermarking strategies bring various improvements: \cite{takezawa2023necessarysufficientwatermarklarge} enhance logit-perturbation, while \cite{hu2023unbiasedwatermarklargelanguage,zhao2024permuteandflipoptimallyrobustwatermarkable} optimize sampling strategies.
Additionally, ~\cite{lee2024wrotecodewatermarkingcode, li2023tracingcodeprovenancecode, yang2021tracingtextprovenancecontextaware} explore code watermarking.

Recent advancements have introduced parameter-based watermarking, which embeds watermarks through distillation~\citep{gu2024learnabilitywatermarkslanguagemodels}.
Other studies focus on investigating typical watermarking behaviors~\citep{luo2024lostoverlapexploringwatermark, singh2023newevaluationmetricscapture}, and some establish robust statistical frameworks for watermarking~\citep{huang2024optimalstatisticalwatermarking, li2024statisticalframeworkwatermarkslarge}. 
Surveys provide detailed definitions and classifications of text watermarking techniques~\citep{jawahar2020automaticdetectionmachinegenerated, liu2024surveytextwatermarkingera,cai2024betterstatisticalunderstandingwatermarking}, while benchmarks offer comprehensive evaluations of watermarks~\citep{tu2024waterbenchholisticevaluationwatermarks}.

\paragraph{Model interventions.}
Beyond fine-tuning, researchers have explored parameter-level interventions to modify model behaviors. 
 Key approaches include model patching~\citep{goel2020modelpatchingclosingsubgroup,ilharco2022patchingopenvocabularymodelsinterpolating,murty2022fixingmodelbugsnatural, sung2021trainingneuralnetworksfixed}, parameter editing~\citep{mitchell2022fastmodeleditingscale, mitchell2022memorybasedmodeleditingscale, santurkar2021editingclassifierrewritingprediction, ilharco2023editingmodelstaskarithmetic}, and model alignment~\citep{askell2021generallanguageassistantlaboratory, glaese2022improvingalignmentdialogueagents, kasirzadeh2022conversationartificialintelligencealigning}. 
Compared to retraining or fine-tuning, model intervention offers a more efficient way to introduce new capabilities into models.

\section{Conclusion}



In this paper, we propose \Ours{}, a training-free, parameter-based watermarking scheme designed for fine-tuned open-source models. 
We evaluate its effectiveness on various model architectures and watermarking strategies. 
Our method resolves the key technical challenges of applying watermarks to fine-tuned models while retaining the fine-tuned model abilities.
Furthermore, we analyze the relationship between parameter integration and the model performance, using cosine similarity analysis to demonstrate that the watermarking parameters encode $n$-gram related knowledge.

Future work could further enhance \Ours{} by developing watermarking strategies better suited to watermark transfer or optimizing the watermark distillation process to produce better watermark-distilled base models.
Additionally, refining the extraction procedure for watermark parameters could improve the efficiency of watermark transfer. 
This would also minimize interference with other model parameters, helping to preserve the overall model performance.

\begin{acks}
To Robert, for the bagels and explaining CMYK and color spaces.
\end{acks}

\bibliographystyle{ACM-Reference-Format}
\bibliography{reference}


\begin{thebibliography}{66}


\ifx \showCODEN    \undefined \def \showCODEN     #1{\unskip}     \fi
\ifx \showDOI      \undefined \def \showDOI       #1{#1}\fi
\ifx \showISBNx    \undefined \def \showISBNx     #1{\unskip}     \fi
\ifx \showISBNxiii \undefined \def \showISBNxiii  #1{\unskip}     \fi
\ifx \showISSN     \undefined \def \showISSN      #1{\unskip}     \fi
\ifx \showLCCN     \undefined \def \showLCCN      #1{\unskip}     \fi
\ifx \shownote     \undefined \def \shownote      #1{#1}          \fi
\ifx \showarticletitle \undefined \def \showarticletitle #1{#1}   \fi
\ifx \showURL      \undefined \def \showURL       {\relax}        \fi
\providecommand\bibfield[2]{#2}
\providecommand\bibinfo[2]{#2}
\providecommand\natexlab[1]{#1}
\providecommand\showeprint[2][]{arXiv:#2}

\bibitem[Aaronson(2023)]%
        {aaronson2023watermarking}
\bibfield{author}{\bibinfo{person}{Scott Aaronson}.} \bibinfo{year}{2023}\natexlab{}.
\newblock \bibinfo{title}{Watermarking of Large Language Models}.
\newblock \bibinfo{howpublished}{Large Language Models and Transformers Workshop at Simons Institute for the Theory of Computing}.
\newblock
\urldef\tempurl%
\url{https://www.youtube.com/watch?v=2Kx9jbSMZqA}
\showURL{%
\tempurl}
\newblock
\shownote{Accessed: September 16, 2024}.


\bibitem[Abdelnabi and Fritz(2021)]%
        {abdelnabi2021adversarialwatermarkingtransformertracing}
\bibfield{author}{\bibinfo{person}{Sahar Abdelnabi} {and} \bibinfo{person}{Mario Fritz}.} \bibinfo{year}{2021}\natexlab{}.
\newblock \bibinfo{title}{Adversarial Watermarking Transformer: Towards Tracing Text Provenance with Data Hiding}.
\newblock
\newblock
\showeprint[arxiv]{2009.03015}~[cs.CR]
\urldef\tempurl%
\url{https://arxiv.org/abs/2009.03015}
\showURL{%
\tempurl}


\bibitem[Agarwalla et~al\mbox{.}(2024)]%
        {agarwalla2024enablinghighsparsityfoundationalllama}
\bibfield{author}{\bibinfo{person}{Abhinav Agarwalla}, \bibinfo{person}{Abhay Gupta}, \bibinfo{person}{Alexandre Marques}, \bibinfo{person}{Shubhra Pandit}, \bibinfo{person}{Michael Goin}, \bibinfo{person}{Eldar Kurtic}, \bibinfo{person}{Kevin Leong}, \bibinfo{person}{Tuan Nguyen}, \bibinfo{person}{Mahmoud Salem}, \bibinfo{person}{Dan Alistarh}, \bibinfo{person}{Sean Lie}, {and} \bibinfo{person}{Mark Kurtz}.} \bibinfo{year}{2024}\natexlab{}.
\newblock \bibinfo{title}{Enabling High-Sparsity Foundational Llama Models with Efficient Pretraining and Deployment}.
\newblock
\newblock
\showeprint[arxiv]{2405.03594}~[cs.CL]
\urldef\tempurl%
\url{https://arxiv.org/abs/2405.03594}
\showURL{%
\tempurl}


\bibitem[Askell et~al\mbox{.}(2021)]%
        {askell2021generallanguageassistantlaboratory}
\bibfield{author}{\bibinfo{person}{Amanda Askell}, \bibinfo{person}{Yuntao Bai}, \bibinfo{person}{Anna Chen}, \bibinfo{person}{Dawn Drain}, \bibinfo{person}{Deep Ganguli}, \bibinfo{person}{Tom Henighan}, \bibinfo{person}{Andy Jones}, \bibinfo{person}{Nicholas Joseph}, \bibinfo{person}{Ben Mann}, \bibinfo{person}{Nova DasSarma}, \bibinfo{person}{Nelson Elhage}, \bibinfo{person}{Zac Hatfield-Dodds}, \bibinfo{person}{Danny Hernandez}, \bibinfo{person}{Jackson Kernion}, \bibinfo{person}{Kamal Ndousse}, \bibinfo{person}{Catherine Olsson}, \bibinfo{person}{Dario Amodei}, \bibinfo{person}{Tom Brown}, \bibinfo{person}{Jack Clark}, \bibinfo{person}{Sam McCandlish}, \bibinfo{person}{Chris Olah}, {and} \bibinfo{person}{Jared Kaplan}.} \bibinfo{year}{2021}\natexlab{}.
\newblock \bibinfo{title}{A General Language Assistant as a Laboratory for Alignment}.
\newblock
\newblock
\showeprint[arxiv]{2112.00861}~[cs.CL]
\urldef\tempurl%
\url{https://arxiv.org/abs/2112.00861}
\showURL{%
\tempurl}


\bibitem[Biderman et~al\mbox{.}(2023)]%
        {biderman2023pythiasuiteanalyzinglarge}
\bibfield{author}{\bibinfo{person}{Stella Biderman}, \bibinfo{person}{Hailey Schoelkopf}, \bibinfo{person}{Quentin Anthony}, \bibinfo{person}{Herbie Bradley}, \bibinfo{person}{Kyle O'Brien}, \bibinfo{person}{Eric Hallahan}, \bibinfo{person}{Mohammad~Aflah Khan}, \bibinfo{person}{Shivanshu Purohit}, \bibinfo{person}{USVSN~Sai Prashanth}, \bibinfo{person}{Edward Raff}, \bibinfo{person}{Aviya Skowron}, \bibinfo{person}{Lintang Sutawika}, {and} \bibinfo{person}{Oskar van~der Wal}.} \bibinfo{year}{2023}\natexlab{}.
\newblock \bibinfo{title}{Pythia: A Suite for Analyzing Large Language Models Across Training and Scaling}.
\newblock
\newblock
\showeprint[arxiv]{2304.01373}~[cs.CL]
\urldef\tempurl%
\url{https://arxiv.org/abs/2304.01373}
\showURL{%
\tempurl}


\bibitem[Cai et~al\mbox{.}(2024)]%
        {cai2024betterstatisticalunderstandingwatermarking}
\bibfield{author}{\bibinfo{person}{Zhongze Cai}, \bibinfo{person}{Shang Liu}, \bibinfo{person}{Hanzhao Wang}, \bibinfo{person}{Huaiyang Zhong}, {and} \bibinfo{person}{Xiaocheng Li}.} \bibinfo{year}{2024}\natexlab{}.
\newblock \bibinfo{title}{Towards Better Statistical Understanding of Watermarking LLMs}.
\newblock
\newblock
\showeprint[arxiv]{2403.13027}~[cs.LG]
\urldef\tempurl%
\url{https://arxiv.org/abs/2403.13027}
\showURL{%
\tempurl}


\bibitem[Cane and Luce(1960)]%
        {Cane1960IndividualCB}
\bibfield{author}{\bibinfo{person}{Violet~R. Cane} {and} \bibinfo{person}{R.~Duncan Luce}.} \bibinfo{year}{1960}\natexlab{}.
\newblock \showarticletitle{Individual Choice Behavior: A Theoretical Analysis.}
\newblock
\urldef\tempurl%
\url{https://api.semanticscholar.org/CorpusID:125306131}
\showURL{%
\tempurl}


\bibitem[Christ et~al\mbox{.}(2023)]%
        {christ2023undetectablewatermarkslanguagemodels}
\bibfield{author}{\bibinfo{person}{Miranda Christ}, \bibinfo{person}{Sam Gunn}, {and} \bibinfo{person}{Or Zamir}.} \bibinfo{year}{2023}\natexlab{}.
\newblock \bibinfo{title}{Undetectable Watermarks for Language Models}.
\newblock
\newblock
\showeprint[arxiv]{2306.09194}~[cs.CR]
\urldef\tempurl%
\url{https://arxiv.org/abs/2306.09194}
\showURL{%
\tempurl}


\bibitem[Cobbe et~al\mbox{.}(2021)]%
        {cobbe2021gsm8kmath}
\bibfield{author}{\bibinfo{person}{Karl Cobbe}, \bibinfo{person}{Vineet Kosaraju}, \bibinfo{person}{Mohammad Bavarian}, \bibinfo{person}{Mark Chen}, \bibinfo{person}{Heewoo Jun}, \bibinfo{person}{Lukasz Kaiser}, \bibinfo{person}{Matthias Plappert}, \bibinfo{person}{Jerry Tworek}, \bibinfo{person}{Jacob Hilton}, \bibinfo{person}{Reiichiro Nakano}, \bibinfo{person}{Christopher Hesse}, {and} \bibinfo{person}{John Schulman}.} \bibinfo{year}{2021}\natexlab{}.
\newblock \bibinfo{title}{Training Verifiers to Solve Math Word Problems}.
\newblock
\newblock
\showeprint[arxiv]{2110.14168}~[cs.LG]
\urldef\tempurl%
\url{https://arxiv.org/abs/2110.14168}
\showURL{%
\tempurl}


\bibitem[Dai and Cai(2019)]%
        {dai-cai-2019-towards}
\bibfield{author}{\bibinfo{person}{Falcon Dai} {and} \bibinfo{person}{Zheng Cai}.} \bibinfo{year}{2019}\natexlab{}.
\newblock \showarticletitle{Towards Near-imperceptible Steganographic Text}. In \bibinfo{booktitle}{\emph{Proceedings of the 57th Annual Meeting of the Association for Computational Linguistics}}, \bibfield{editor}{\bibinfo{person}{Anna Korhonen}, \bibinfo{person}{David Traum}, {and} \bibinfo{person}{Llu{\'\i}s M{\`a}rquez}} (Eds.). \bibinfo{publisher}{Association for Computational Linguistics}, \bibinfo{address}{Florence, Italy}, \bibinfo{pages}{4303--4308}.
\newblock
\urldef\tempurl%
\url{https://doi.org/10.18653/v1/P19-1422}
\showDOI{\tempurl}


\bibitem[Dai et~al\mbox{.}(2024)]%
        {safe-rlhf}
\bibfield{author}{\bibinfo{person}{Josef Dai}, \bibinfo{person}{Xuehai Pan}, \bibinfo{person}{Ruiyang Sun}, \bibinfo{person}{Jiaming Ji}, \bibinfo{person}{Xinbo Xu}, \bibinfo{person}{Mickel Liu}, \bibinfo{person}{Yizhou Wang}, {and} \bibinfo{person}{Yaodong Yang}.} \bibinfo{year}{2024}\natexlab{}.
\newblock \showarticletitle{Safe RLHF: Safe Reinforcement Learning from Human Feedback}. In \bibinfo{booktitle}{\emph{The Twelfth International Conference on Learning Representations}}.
\newblock
\urldef\tempurl%
\url{https://openreview.net/forum?id=TyFrPOKYXw}
\showURL{%
\tempurl}


\bibitem[Fang et~al\mbox{.}(2017)]%
        {fang-etal-2017-generating}
\bibfield{author}{\bibinfo{person}{Tina Fang}, \bibinfo{person}{Martin Jaggi}, {and} \bibinfo{person}{Katerina Argyraki}.} \bibinfo{year}{2017}\natexlab{}.
\newblock \showarticletitle{Generating Steganographic Text with {LSTM}s}. In \bibinfo{booktitle}{\emph{Proceedings of {ACL} 2017, Student Research Workshop}}, \bibfield{editor}{\bibinfo{person}{Allyson Ettinger}, \bibinfo{person}{Spandana Gella}, \bibinfo{person}{Matthieu Labeau}, \bibinfo{person}{Cecilia~Ovesdotter Alm}, \bibinfo{person}{Marine Carpuat}, {and} \bibinfo{person}{Mark Dredze}} (Eds.). \bibinfo{publisher}{Association for Computational Linguistics}, \bibinfo{address}{Vancouver, Canada}, \bibinfo{pages}{100--106}.
\newblock
\urldef\tempurl%
\url{https://aclanthology.org/P17-3017}
\showURL{%
\tempurl}


\bibitem[Giboulot and Teddy(2024)]%
        {giboulot2024watermaxbreakingllmwatermark}
\bibfield{author}{\bibinfo{person}{Eva Giboulot} {and} \bibinfo{person}{Furon Teddy}.} \bibinfo{year}{2024}\natexlab{}.
\newblock \bibinfo{title}{WaterMax: breaking the LLM watermark detectability-robustness-quality trade-off}.
\newblock
\newblock
\showeprint[arxiv]{2403.04808}~[cs.CR]
\urldef\tempurl%
\url{https://arxiv.org/abs/2403.04808}
\showURL{%
\tempurl}


\bibitem[Glaese et~al\mbox{.}(2022)]%
        {glaese2022improvingalignmentdialogueagents}
\bibfield{author}{\bibinfo{person}{Amelia Glaese}, \bibinfo{person}{Nat McAleese}, \bibinfo{person}{Maja Trebacz}, \bibinfo{person}{John Aslanides}, \bibinfo{person}{Vlad Firoiu}, \bibinfo{person}{Timo Ewalds}, \bibinfo{person}{Maribeth Rauh}, \bibinfo{person}{Laura Weidinger}, \bibinfo{person}{Martin Chadwick}, \bibinfo{person}{Phoebe Thacker}, \bibinfo{person}{Lucy Campbell-Gillingham}, \bibinfo{person}{Jonathan Uesato}, \bibinfo{person}{Po-Sen Huang}, \bibinfo{person}{Ramona Comanescu}, \bibinfo{person}{Fan Yang}, \bibinfo{person}{Abigail See}, \bibinfo{person}{Sumanth Dathathri}, \bibinfo{person}{Rory Greig}, \bibinfo{person}{Charlie Chen}, \bibinfo{person}{Doug Fritz}, \bibinfo{person}{Jaume~Sanchez Elias}, \bibinfo{person}{Richard Green}, \bibinfo{person}{Sona Mokra}, \bibinfo{person}{Nicholas Fernando}, \bibinfo{person}{Boxi Wu}, \bibinfo{person}{Rachel Foley}, \bibinfo{person}{Susannah Young}, \bibinfo{person}{Iason Gabriel}, \bibinfo{person}{William Isaac}, \bibinfo{person}{John Mellor},
  \bibinfo{person}{Demis Hassabis}, \bibinfo{person}{Koray Kavukcuoglu}, \bibinfo{person}{Lisa~Anne Hendricks}, {and} \bibinfo{person}{Geoffrey Irving}.} \bibinfo{year}{2022}\natexlab{}.
\newblock \bibinfo{title}{Improving alignment of dialogue agents via targeted human judgements}.
\newblock
\newblock
\showeprint[arxiv]{2209.14375}~[cs.LG]
\urldef\tempurl%
\url{https://arxiv.org/abs/2209.14375}
\showURL{%
\tempurl}


\bibitem[Goel et~al\mbox{.}(2020)]%
        {goel2020modelpatchingclosingsubgroup}
\bibfield{author}{\bibinfo{person}{Karan Goel}, \bibinfo{person}{Albert Gu}, \bibinfo{person}{Yixuan Li}, {and} \bibinfo{person}{Christopher Ré}.} \bibinfo{year}{2020}\natexlab{}.
\newblock \bibinfo{title}{Model Patching: Closing the Subgroup Performance Gap with Data Augmentation}.
\newblock
\newblock
\showeprint[arxiv]{2008.06775}~[cs.LG]
\urldef\tempurl%
\url{https://arxiv.org/abs/2008.06775}
\showURL{%
\tempurl}


\bibitem[Gokaslan and Cohen(2019)]%
        {Gokaslan2019OpenWeb}
\bibfield{author}{\bibinfo{person}{Aaron Gokaslan} {and} \bibinfo{person}{Vanya Cohen}.} \bibinfo{year}{2019}\natexlab{}.
\newblock \bibinfo{title}{OpenWebText Corpus}.
\newblock \bibinfo{howpublished}{\url{http://Skylion007.github.io/OpenWebTextCorpus}}.
\newblock


\bibitem[Gu et~al\mbox{.}(2024)]%
        {gu2024learnabilitywatermarkslanguagemodels}
\bibfield{author}{\bibinfo{person}{Chenchen Gu}, \bibinfo{person}{Xiang~Lisa Li}, \bibinfo{person}{Percy Liang}, {and} \bibinfo{person}{Tatsunori Hashimoto}.} \bibinfo{year}{2024}\natexlab{}.
\newblock \bibinfo{title}{On the Learnability of Watermarks for Language Models}.
\newblock
\newblock
\showeprint[arxiv]{2312.04469}~[cs.LG]
\urldef\tempurl%
\url{https://arxiv.org/abs/2312.04469}
\showURL{%
\tempurl}


\bibitem[Hendrycks et~al\mbox{.}(2021)]%
        {hendrycks2021measuringmassivemultitasklanguage}
\bibfield{author}{\bibinfo{person}{Dan Hendrycks}, \bibinfo{person}{Collin Burns}, \bibinfo{person}{Steven Basart}, \bibinfo{person}{Andy Zou}, \bibinfo{person}{Mantas Mazeika}, \bibinfo{person}{Dawn Song}, {and} \bibinfo{person}{Jacob Steinhardt}.} \bibinfo{year}{2021}\natexlab{}.
\newblock \bibinfo{title}{Measuring Massive Multitask Language Understanding}.
\newblock
\newblock
\showeprint[arxiv]{2009.03300}~[cs.CY]
\urldef\tempurl%
\url{https://arxiv.org/abs/2009.03300}
\showURL{%
\tempurl}


\bibitem[Hu et~al\mbox{.}(2023)]%
        {hu2023unbiasedwatermarklargelanguage}
\bibfield{author}{\bibinfo{person}{Zhengmian Hu}, \bibinfo{person}{Lichang Chen}, \bibinfo{person}{Xidong Wu}, \bibinfo{person}{Yihan Wu}, \bibinfo{person}{Hongyang Zhang}, {and} \bibinfo{person}{Heng Huang}.} \bibinfo{year}{2023}\natexlab{}.
\newblock \bibinfo{title}{Unbiased Watermark for Large Language Models}.
\newblock
\newblock
\showeprint[arxiv]{2310.10669}~[cs.CR]
\urldef\tempurl%
\url{https://arxiv.org/abs/2310.10669}
\showURL{%
\tempurl}


\bibitem[Huang et~al\mbox{.}(2024)]%
        {huang2024optimalstatisticalwatermarking}
\bibfield{author}{\bibinfo{person}{Baihe Huang}, \bibinfo{person}{Hanlin Zhu}, \bibinfo{person}{Banghua Zhu}, \bibinfo{person}{Kannan Ramchandran}, \bibinfo{person}{Michael~I. Jordan}, \bibinfo{person}{Jason~D. Lee}, {and} \bibinfo{person}{Jiantao Jiao}.} \bibinfo{year}{2024}\natexlab{}.
\newblock \bibinfo{title}{Towards Optimal Statistical Watermarking}.
\newblock
\newblock
\showeprint[arxiv]{2312.07930}~[cs.LG]
\urldef\tempurl%
\url{https://arxiv.org/abs/2312.07930}
\showURL{%
\tempurl}


\bibitem[Ilharco et~al\mbox{.}(2023)]%
        {ilharco2023editingmodelstaskarithmetic}
\bibfield{author}{\bibinfo{person}{Gabriel Ilharco}, \bibinfo{person}{Marco~Tulio Ribeiro}, \bibinfo{person}{Mitchell Wortsman}, \bibinfo{person}{Suchin Gururangan}, \bibinfo{person}{Ludwig Schmidt}, \bibinfo{person}{Hannaneh Hajishirzi}, {and} \bibinfo{person}{Ali Farhadi}.} \bibinfo{year}{2023}\natexlab{}.
\newblock \bibinfo{title}{Editing Models with Task Arithmetic}.
\newblock
\newblock
\showeprint[arxiv]{2212.04089}~[cs.LG]
\urldef\tempurl%
\url{https://arxiv.org/abs/2212.04089}
\showURL{%
\tempurl}


\bibitem[Ilharco et~al\mbox{.}(2022)]%
        {ilharco2022patchingopenvocabularymodelsinterpolating}
\bibfield{author}{\bibinfo{person}{Gabriel Ilharco}, \bibinfo{person}{Mitchell Wortsman}, \bibinfo{person}{Samir~Yitzhak Gadre}, \bibinfo{person}{Shuran Song}, \bibinfo{person}{Hannaneh Hajishirzi}, \bibinfo{person}{Simon Kornblith}, \bibinfo{person}{Ali Farhadi}, {and} \bibinfo{person}{Ludwig Schmidt}.} \bibinfo{year}{2022}\natexlab{}.
\newblock \bibinfo{title}{Patching open-vocabulary models by interpolating weights}.
\newblock
\newblock
\showeprint[arxiv]{2208.05592}~[cs.CV]
\urldef\tempurl%
\url{https://arxiv.org/abs/2208.05592}
\showURL{%
\tempurl}


\bibitem[Jawahar et~al\mbox{.}(2020)]%
        {jawahar2020automaticdetectionmachinegenerated}
\bibfield{author}{\bibinfo{person}{Ganesh Jawahar}, \bibinfo{person}{Muhammad Abdul-Mageed}, {and} \bibinfo{person}{Laks V.~S. Lakshmanan}.} \bibinfo{year}{2020}\natexlab{}.
\newblock \bibinfo{title}{Automatic Detection of Machine Generated Text: A Critical Survey}.
\newblock
\newblock
\showeprint[arxiv]{2011.01314}~[cs.CL]
\urldef\tempurl%
\url{https://arxiv.org/abs/2011.01314}
\showURL{%
\tempurl}


\bibitem[Jiao et~al\mbox{.}(2024)]%
        {jiao2024incontextprobingapproximatesinfluence}
\bibfield{author}{\bibinfo{person}{Cathy Jiao}, \bibinfo{person}{Gary Gao}, {and} \bibinfo{person}{Chenyan Xiong}.} \bibinfo{year}{2024}\natexlab{}.
\newblock \bibinfo{title}{In-Context Probing Approximates Influence Function for Data Valuation}.
\newblock
\newblock
\showeprint[arxiv]{2407.12259}~[cs.CL]
\urldef\tempurl%
\url{https://arxiv.org/abs/2407.12259}
\showURL{%
\tempurl}


\bibitem[Kasirzadeh and Gabriel(2022)]%
        {kasirzadeh2022conversationartificialintelligencealigning}
\bibfield{author}{\bibinfo{person}{Atoosa Kasirzadeh} {and} \bibinfo{person}{Iason Gabriel}.} \bibinfo{year}{2022}\natexlab{}.
\newblock \bibinfo{title}{In conversation with Artificial Intelligence: aligning language models with human values}.
\newblock
\newblock
\showeprint[arxiv]{2209.00731}~[cs.CY]
\urldef\tempurl%
\url{https://arxiv.org/abs/2209.00731}
\showURL{%
\tempurl}


\bibitem[Kirchenbauer et~al\mbox{.}(2024a)]%
        {kirchenbauer2024watermarklargelanguagemodels}
\bibfield{author}{\bibinfo{person}{John Kirchenbauer}, \bibinfo{person}{Jonas Geiping}, \bibinfo{person}{Yuxin Wen}, \bibinfo{person}{Jonathan Katz}, \bibinfo{person}{Ian Miers}, {and} \bibinfo{person}{Tom Goldstein}.} \bibinfo{year}{2024}\natexlab{a}.
\newblock \bibinfo{title}{A Watermark for Large Language Models}.
\newblock
\newblock
\showeprint[arxiv]{2301.10226}~[cs.LG]
\urldef\tempurl%
\url{https://arxiv.org/abs/2301.10226}
\showURL{%
\tempurl}


\bibitem[Kirchenbauer et~al\mbox{.}(2024b)]%
        {kirchenbauer2024reliabilitywatermarkslargelanguage}
\bibfield{author}{\bibinfo{person}{John Kirchenbauer}, \bibinfo{person}{Jonas Geiping}, \bibinfo{person}{Yuxin Wen}, \bibinfo{person}{Manli Shu}, \bibinfo{person}{Khalid Saifullah}, \bibinfo{person}{Kezhi Kong}, \bibinfo{person}{Kasun Fernando}, \bibinfo{person}{Aniruddha Saha}, \bibinfo{person}{Micah Goldblum}, {and} \bibinfo{person}{Tom Goldstein}.} \bibinfo{year}{2024}\natexlab{b}.
\newblock \bibinfo{title}{On the Reliability of Watermarks for Large Language Models}.
\newblock
\newblock
\showeprint[arxiv]{2306.04634}~[cs.LG]
\urldef\tempurl%
\url{https://arxiv.org/abs/2306.04634}
\showURL{%
\tempurl}


\bibitem[Krishna et~al\mbox{.}(2023)]%
        {krishnaparaphraseattack2023}
\bibfield{author}{\bibinfo{person}{Kalpesh Krishna}, \bibinfo{person}{Yixiao Song}, \bibinfo{person}{Marzena Karpinska}, \bibinfo{person}{John Wieting}, {and} \bibinfo{person}{Mohit Iyyer}.} \bibinfo{year}{2023}\natexlab{}.
\newblock \showarticletitle{Paraphrasing evades detectors of AI-generated text, but retrieval is an effective defense}. In \bibinfo{booktitle}{\emph{Advances in Neural Information Processing Systems 36: Annual Conference on Neural Information Processing Systems 2023, NeurIPS 2023, New Orleans, LA, USA, December 10 - 16, 2023}}, \bibfield{editor}{\bibinfo{person}{Alice Oh}, \bibinfo{person}{Tristan Naumann}, \bibinfo{person}{Amir Globerson}, \bibinfo{person}{Kate Saenko}, \bibinfo{person}{Moritz Hardt}, {and} \bibinfo{person}{Sergey Levine}} (Eds.).
\newblock
\urldef\tempurl%
\url{http://papers.nips.cc/paper\_files/paper/2023/hash/575c450013d0e99e4b0ecf82bd1afaa4-Abstract-Conference.html}
\showURL{%
\tempurl}


\bibitem[Kuditipudi et~al\mbox{.}(2024)]%
        {kuditipudi2024robustdistortionfreewatermarkslanguage}
\bibfield{author}{\bibinfo{person}{Rohith Kuditipudi}, \bibinfo{person}{John Thickstun}, \bibinfo{person}{Tatsunori Hashimoto}, {and} \bibinfo{person}{Percy Liang}.} \bibinfo{year}{2024}\natexlab{}.
\newblock \bibinfo{title}{Robust Distortion-free Watermarks for Language Models}.
\newblock
\newblock
\showeprint[arxiv]{2307.15593}~[cs.LG]
\urldef\tempurl%
\url{https://arxiv.org/abs/2307.15593}
\showURL{%
\tempurl}


\bibitem[Labs(2024)]%
        {lambdalabs_pythia_1.4b_2024}
\bibfield{author}{\bibinfo{person}{Lambda Labs}.} \bibinfo{year}{2024}\natexlab{}.
\newblock \bibinfo{title}{Pythia 1.4B Deduped Synthetic Instruct Model}.
\newblock
\newblock
\urldef\tempurl%
\url{https://huggingface.co/lambdalabs/pythia-1.4b-deduped-synthetic-instruct}
\showURL{%
\tempurl}
\newblock
\shownote{Accessed: 7/31/2024}.


\bibitem[Lee et~al\mbox{.}(2024)]%
        {lee2024wrotecodewatermarkingcode}
\bibfield{author}{\bibinfo{person}{Taehyun Lee}, \bibinfo{person}{Seokhee Hong}, \bibinfo{person}{Jaewoo Ahn}, \bibinfo{person}{Ilgee Hong}, \bibinfo{person}{Hwaran Lee}, \bibinfo{person}{Sangdoo Yun}, \bibinfo{person}{Jamin Shin}, {and} \bibinfo{person}{Gunhee Kim}.} \bibinfo{year}{2024}\natexlab{}.
\newblock \bibinfo{title}{Who Wrote this Code? Watermarking for Code Generation}.
\newblock
\newblock
\showeprint[arxiv]{2305.15060}~[cs.CL]
\urldef\tempurl%
\url{https://arxiv.org/abs/2305.15060}
\showURL{%
\tempurl}


\bibitem[Li et~al\mbox{.}(2023)]%
        {li2023tracingcodeprovenancecode}
\bibfield{author}{\bibinfo{person}{Wei Li}, \bibinfo{person}{Borui Yang}, \bibinfo{person}{Yujie Sun}, \bibinfo{person}{Suyu Chen}, \bibinfo{person}{Ziyun Song}, \bibinfo{person}{Liyao Xiang}, \bibinfo{person}{Xinbing Wang}, {and} \bibinfo{person}{Chenghu Zhou}.} \bibinfo{year}{2023}\natexlab{}.
\newblock \bibinfo{title}{Towards Tracing Code Provenance with Code Watermarking}.
\newblock
\newblock
\showeprint[arxiv]{2305.12461}~[cs.CR]
\urldef\tempurl%
\url{https://arxiv.org/abs/2305.12461}
\showURL{%
\tempurl}


\bibitem[Li et~al\mbox{.}(2024)]%
        {li2024statisticalframeworkwatermarkslarge}
\bibfield{author}{\bibinfo{person}{Xiang Li}, \bibinfo{person}{Feng Ruan}, \bibinfo{person}{Huiyuan Wang}, \bibinfo{person}{Qi Long}, {and} \bibinfo{person}{Weijie~J. Su}.} \bibinfo{year}{2024}\natexlab{}.
\newblock \bibinfo{title}{A Statistical Framework of Watermarks for Large Language Models: Pivot, Detection Efficiency and Optimal Rules}.
\newblock
\newblock
\showeprint[arxiv]{2404.01245}~[math.ST]
\urldef\tempurl%
\url{https://arxiv.org/abs/2404.01245}
\showURL{%
\tempurl}


\bibitem[Lin(1991)]%
        {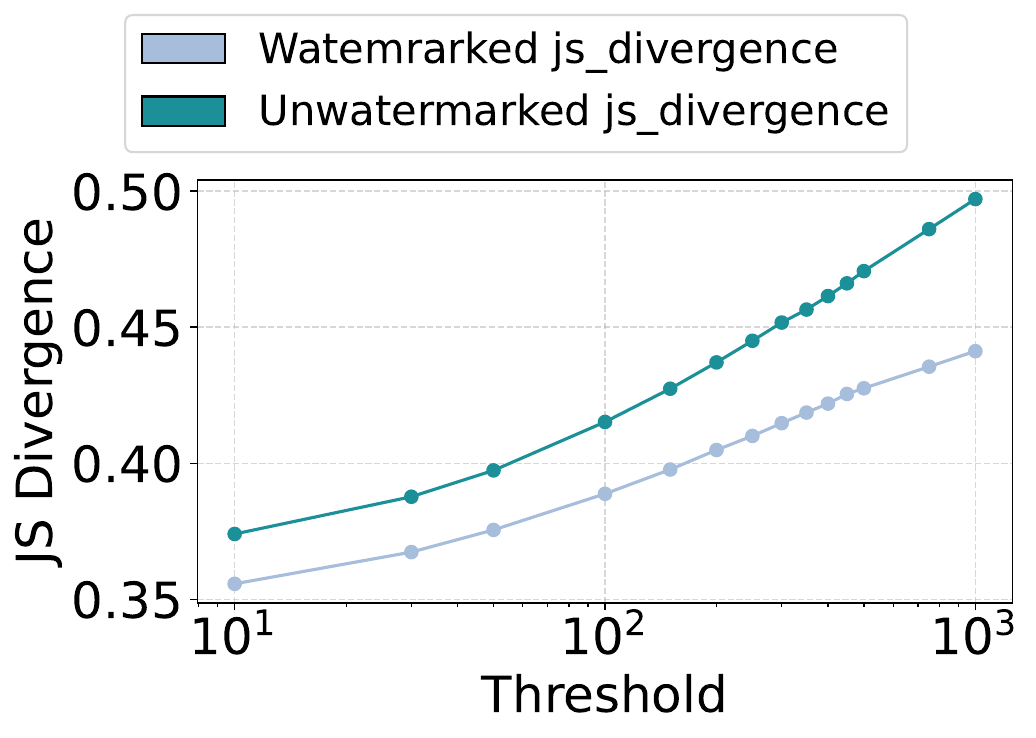}
\bibfield{author}{\bibinfo{person}{J. Lin}.} \bibinfo{year}{1991}\natexlab{}.
\newblock \showarticletitle{Divergence measures based on the Shannon entropy}.
\newblock \bibinfo{journal}{\emph{IEEE Transactions on Information Theory}} \bibinfo{volume}{37}, \bibinfo{number}{1} (\bibinfo{year}{1991}), \bibinfo{pages}{145--151}.
\newblock
\urldef\tempurl%
\url{https://doi.org/10.1109/18.61115}
\showDOI{\tempurl}


\bibitem[Liu et~al\mbox{.}(2024)]%
        {liu2024surveytextwatermarkingera}
\bibfield{author}{\bibinfo{person}{Aiwei Liu}, \bibinfo{person}{Leyi Pan}, \bibinfo{person}{Yijian Lu}, \bibinfo{person}{Jingjing Li}, \bibinfo{person}{Xuming Hu}, \bibinfo{person}{Xi Zhang}, \bibinfo{person}{Lijie Wen}, \bibinfo{person}{Irwin King}, \bibinfo{person}{Hui Xiong}, {and} \bibinfo{person}{Philip~S. Yu}.} \bibinfo{year}{2024}\natexlab{}.
\newblock \bibinfo{title}{A Survey of Text Watermarking in the Era of Large Language Models}.
\newblock
\newblock
\showeprint[arxiv]{2312.07913}~[cs.CL]
\urldef\tempurl%
\url{https://arxiv.org/abs/2312.07913}
\showURL{%
\tempurl}


\bibitem[Liu and Bu(2024)]%
        {liu2024adaptivetextwatermarklarge}
\bibfield{author}{\bibinfo{person}{Yepeng Liu} {and} \bibinfo{person}{Yuheng Bu}.} \bibinfo{year}{2024}\natexlab{}.
\newblock \bibinfo{title}{Adaptive Text Watermark for Large Language Models}.
\newblock
\newblock
\showeprint[arxiv]{2401.13927}~[cs.CL]
\urldef\tempurl%
\url{https://arxiv.org/abs/2401.13927}
\showURL{%
\tempurl}


\bibitem[Lu et~al\mbox{.}(2024)]%
        {lu2024entropybasedtextwatermarkingdetection}
\bibfield{author}{\bibinfo{person}{Yijian Lu}, \bibinfo{person}{Aiwei Liu}, \bibinfo{person}{Dianzhi Yu}, \bibinfo{person}{Jingjing Li}, {and} \bibinfo{person}{Irwin King}.} \bibinfo{year}{2024}\natexlab{}.
\newblock \bibinfo{title}{An Entropy-based Text Watermarking Detection Method}.
\newblock
\newblock
\showeprint[arxiv]{2403.13485}~[cs.CL]
\urldef\tempurl%
\url{https://arxiv.org/abs/2403.13485}
\showURL{%
\tempurl}


\bibitem[Luo et~al\mbox{.}(2024)]%
        {luo2024lostoverlapexploringwatermark}
\bibfield{author}{\bibinfo{person}{Yiyang Luo}, \bibinfo{person}{Ke Lin}, {and} \bibinfo{person}{Chao Gu}.} \bibinfo{year}{2024}\natexlab{}.
\newblock \bibinfo{title}{Lost in Overlap: Exploring Watermark Collision in LLMs}.
\newblock
\newblock
\showeprint[arxiv]{2403.10020}~[cs.CL]
\urldef\tempurl%
\url{https://arxiv.org/abs/2403.10020}
\showURL{%
\tempurl}


\bibitem[Mitchell et~al\mbox{.}(2022a)]%
        {mitchell2022fastmodeleditingscale}
\bibfield{author}{\bibinfo{person}{Eric Mitchell}, \bibinfo{person}{Charles Lin}, \bibinfo{person}{Antoine Bosselut}, \bibinfo{person}{Chelsea Finn}, {and} \bibinfo{person}{Christopher~D. Manning}.} \bibinfo{year}{2022}\natexlab{a}.
\newblock \bibinfo{title}{Fast Model Editing at Scale}.
\newblock
\newblock
\showeprint[arxiv]{2110.11309}~[cs.LG]
\urldef\tempurl%
\url{https://arxiv.org/abs/2110.11309}
\showURL{%
\tempurl}


\bibitem[Mitchell et~al\mbox{.}(2022b)]%
        {mitchell2022memorybasedmodeleditingscale}
\bibfield{author}{\bibinfo{person}{Eric Mitchell}, \bibinfo{person}{Charles Lin}, \bibinfo{person}{Antoine Bosselut}, \bibinfo{person}{Christopher~D. Manning}, {and} \bibinfo{person}{Chelsea Finn}.} \bibinfo{year}{2022}\natexlab{b}.
\newblock \bibinfo{title}{Memory-Based Model Editing at Scale}.
\newblock
\newblock
\showeprint[arxiv]{2206.06520}~[cs.AI]
\urldef\tempurl%
\url{https://arxiv.org/abs/2206.06520}
\showURL{%
\tempurl}


\bibitem[Murty et~al\mbox{.}(2022)]%
        {murty2022fixingmodelbugsnatural}
\bibfield{author}{\bibinfo{person}{Shikhar Murty}, \bibinfo{person}{Christopher~D. Manning}, \bibinfo{person}{Scott Lundberg}, {and} \bibinfo{person}{Marco~Tulio Ribeiro}.} \bibinfo{year}{2022}\natexlab{}.
\newblock \bibinfo{title}{Fixing Model Bugs with Natural Language Patches}.
\newblock
\newblock
\showeprint[arxiv]{2211.03318}~[cs.CL]
\urldef\tempurl%
\url{https://arxiv.org/abs/2211.03318}
\showURL{%
\tempurl}


\bibitem[OpenAI et~al\mbox{.}(2024)]%
        {openai2024gpt4technicalreport}
\bibfield{author}{\bibinfo{person}{OpenAI}, \bibinfo{person}{Josh Achiam}, \bibinfo{person}{Steven Adler}, \bibinfo{person}{Sandhini Agarwal}, \bibinfo{person}{Lama Ahmad}, \bibinfo{person}{Ilge Akkaya}, \bibinfo{person}{Florencia~Leoni Aleman}, \bibinfo{person}{Diogo Almeida}, \bibinfo{person}{Janko Altenschmidt}, \bibinfo{person}{Sam Altman}, \bibinfo{person}{Shyamal Anadkat}, \bibinfo{person}{Red Avila}, \bibinfo{person}{Igor Babuschkin}, \bibinfo{person}{Suchir Balaji}, \bibinfo{person}{Valerie Balcom}, \bibinfo{person}{Paul Baltescu}, \bibinfo{person}{Haiming Bao}, \bibinfo{person}{Mohammad Bavarian}, \bibinfo{person}{Jeff Belgum}, \bibinfo{person}{Irwan Bello}, \bibinfo{person}{Jake Berdine}, \bibinfo{person}{Gabriel Bernadett-Shapiro}, \bibinfo{person}{Christopher Berner}, \bibinfo{person}{Lenny Bogdonoff}, \bibinfo{person}{Oleg Boiko}, \bibinfo{person}{Madelaine Boyd}, \bibinfo{person}{Anna-Luisa Brakman}, \bibinfo{person}{Greg Brockman}, \bibinfo{person}{Tim Brooks}, \bibinfo{person}{Miles Brundage},
  \bibinfo{person}{Kevin Button}, \bibinfo{person}{Trevor Cai}, \bibinfo{person}{Rosie Campbell}, \bibinfo{person}{Andrew Cann}, \bibinfo{person}{Brittany Carey}, \bibinfo{person}{Chelsea Carlson}, \bibinfo{person}{Rory Carmichael}, \bibinfo{person}{Brooke Chan}, \bibinfo{person}{Che Chang}, \bibinfo{person}{Fotis Chantzis}, \bibinfo{person}{Derek Chen}, \bibinfo{person}{Sully Chen}, \bibinfo{person}{Ruby Chen}, \bibinfo{person}{Jason Chen}, \bibinfo{person}{Mark Chen}, \bibinfo{person}{Ben Chess}, \bibinfo{person}{Chester Cho}, \bibinfo{person}{Casey Chu}, \bibinfo{person}{Hyung~Won Chung}, \bibinfo{person}{Dave Cummings}, \bibinfo{person}{Jeremiah Currier}, \bibinfo{person}{Yunxing Dai}, \bibinfo{person}{Cory Decareaux}, \bibinfo{person}{Thomas Degry}, \bibinfo{person}{Noah Deutsch}, \bibinfo{person}{Damien Deville}, \bibinfo{person}{Arka Dhar}, \bibinfo{person}{David Dohan}, \bibinfo{person}{Steve Dowling}, \bibinfo{person}{Sheila Dunning}, \bibinfo{person}{Adrien Ecoffet}, \bibinfo{person}{Atty Eleti},
  \bibinfo{person}{Tyna Eloundou}, \bibinfo{person}{David Farhi}, \bibinfo{person}{Liam Fedus}, \bibinfo{person}{Niko Felix}, \bibinfo{person}{Simón~Posada Fishman}, \bibinfo{person}{Juston Forte}, \bibinfo{person}{Isabella Fulford}, \bibinfo{person}{Leo Gao}, \bibinfo{person}{Elie Georges}, \bibinfo{person}{Christian Gibson}, \bibinfo{person}{Vik Goel}, \bibinfo{person}{Tarun Gogineni}, \bibinfo{person}{Gabriel Goh}, \bibinfo{person}{Rapha Gontijo-Lopes}, \bibinfo{person}{Jonathan Gordon}, \bibinfo{person}{Morgan Grafstein}, \bibinfo{person}{Scott Gray}, \bibinfo{person}{Ryan Greene}, \bibinfo{person}{Joshua Gross}, \bibinfo{person}{Shixiang~Shane Gu}, \bibinfo{person}{Yufei Guo}, \bibinfo{person}{Chris Hallacy}, \bibinfo{person}{Jesse Han}, \bibinfo{person}{Jeff Harris}, \bibinfo{person}{Yuchen He}, \bibinfo{person}{Mike Heaton}, \bibinfo{person}{Johannes Heidecke}, \bibinfo{person}{Chris Hesse}, \bibinfo{person}{Alan Hickey}, \bibinfo{person}{Wade Hickey}, \bibinfo{person}{Peter Hoeschele},
  \bibinfo{person}{Brandon Houghton}, \bibinfo{person}{Kenny Hsu}, \bibinfo{person}{Shengli Hu}, \bibinfo{person}{Xin Hu}, \bibinfo{person}{Joost Huizinga}, \bibinfo{person}{Shantanu Jain}, \bibinfo{person}{Shawn Jain}, \bibinfo{person}{Joanne Jang}, \bibinfo{person}{Angela Jiang}, \bibinfo{person}{Roger Jiang}, \bibinfo{person}{Haozhun Jin}, \bibinfo{person}{Denny Jin}, \bibinfo{person}{Shino Jomoto}, \bibinfo{person}{Billie Jonn}, \bibinfo{person}{Heewoo Jun}, \bibinfo{person}{Tomer Kaftan}, \bibinfo{person}{Łukasz Kaiser}, \bibinfo{person}{Ali Kamali}, \bibinfo{person}{Ingmar Kanitscheider}, \bibinfo{person}{Nitish~Shirish Keskar}, \bibinfo{person}{Tabarak Khan}, \bibinfo{person}{Logan Kilpatrick}, \bibinfo{person}{Jong~Wook Kim}, \bibinfo{person}{Christina Kim}, \bibinfo{person}{Yongjik Kim}, \bibinfo{person}{Jan~Hendrik Kirchner}, \bibinfo{person}{Jamie Kiros}, \bibinfo{person}{Matt Knight}, \bibinfo{person}{Daniel Kokotajlo}, \bibinfo{person}{Łukasz Kondraciuk}, \bibinfo{person}{Andrew Kondrich},
  \bibinfo{person}{Aris Konstantinidis}, \bibinfo{person}{Kyle Kosic}, \bibinfo{person}{Gretchen Krueger}, \bibinfo{person}{Vishal Kuo}, \bibinfo{person}{Michael Lampe}, \bibinfo{person}{Ikai Lan}, \bibinfo{person}{Teddy Lee}, \bibinfo{person}{Jan Leike}, \bibinfo{person}{Jade Leung}, \bibinfo{person}{Daniel Levy}, \bibinfo{person}{Chak~Ming Li}, \bibinfo{person}{Rachel Lim}, \bibinfo{person}{Molly Lin}, \bibinfo{person}{Stephanie Lin}, \bibinfo{person}{Mateusz Litwin}, \bibinfo{person}{Theresa Lopez}, \bibinfo{person}{Ryan Lowe}, \bibinfo{person}{Patricia Lue}, \bibinfo{person}{Anna Makanju}, \bibinfo{person}{Kim Malfacini}, \bibinfo{person}{Sam Manning}, \bibinfo{person}{Todor Markov}, \bibinfo{person}{Yaniv Markovski}, \bibinfo{person}{Bianca Martin}, \bibinfo{person}{Katie Mayer}, \bibinfo{person}{Andrew Mayne}, \bibinfo{person}{Bob McGrew}, \bibinfo{person}{Scott~Mayer McKinney}, \bibinfo{person}{Christine McLeavey}, \bibinfo{person}{Paul McMillan}, \bibinfo{person}{Jake McNeil}, \bibinfo{person}{David
  Medina}, \bibinfo{person}{Aalok Mehta}, \bibinfo{person}{Jacob Menick}, \bibinfo{person}{Luke Metz}, \bibinfo{person}{Andrey Mishchenko}, \bibinfo{person}{Pamela Mishkin}, \bibinfo{person}{Vinnie Monaco}, \bibinfo{person}{Evan Morikawa}, \bibinfo{person}{Daniel Mossing}, \bibinfo{person}{Tong Mu}, \bibinfo{person}{Mira Murati}, \bibinfo{person}{Oleg Murk}, \bibinfo{person}{David Mély}, \bibinfo{person}{Ashvin Nair}, \bibinfo{person}{Reiichiro Nakano}, \bibinfo{person}{Rajeev Nayak}, \bibinfo{person}{Arvind Neelakantan}, \bibinfo{person}{Richard Ngo}, \bibinfo{person}{Hyeonwoo Noh}, \bibinfo{person}{Long Ouyang}, \bibinfo{person}{Cullen O'Keefe}, \bibinfo{person}{Jakub Pachocki}, \bibinfo{person}{Alex Paino}, \bibinfo{person}{Joe Palermo}, \bibinfo{person}{Ashley Pantuliano}, \bibinfo{person}{Giambattista Parascandolo}, \bibinfo{person}{Joel Parish}, \bibinfo{person}{Emy Parparita}, \bibinfo{person}{Alex Passos}, \bibinfo{person}{Mikhail Pavlov}, \bibinfo{person}{Andrew Peng}, \bibinfo{person}{Adam
  Perelman}, \bibinfo{person}{Filipe de Avila Belbute~Peres}, \bibinfo{person}{Michael Petrov}, \bibinfo{person}{Henrique~Ponde de Oliveira~Pinto}, \bibinfo{person}{Michael}, \bibinfo{person}{Pokorny}, \bibinfo{person}{Michelle Pokrass}, \bibinfo{person}{Vitchyr~H. Pong}, \bibinfo{person}{Tolly Powell}, \bibinfo{person}{Alethea Power}, \bibinfo{person}{Boris Power}, \bibinfo{person}{Elizabeth Proehl}, \bibinfo{person}{Raul Puri}, \bibinfo{person}{Alec Radford}, \bibinfo{person}{Jack Rae}, \bibinfo{person}{Aditya Ramesh}, \bibinfo{person}{Cameron Raymond}, \bibinfo{person}{Francis Real}, \bibinfo{person}{Kendra Rimbach}, \bibinfo{person}{Carl Ross}, \bibinfo{person}{Bob Rotsted}, \bibinfo{person}{Henri Roussez}, \bibinfo{person}{Nick Ryder}, \bibinfo{person}{Mario Saltarelli}, \bibinfo{person}{Ted Sanders}, \bibinfo{person}{Shibani Santurkar}, \bibinfo{person}{Girish Sastry}, \bibinfo{person}{Heather Schmidt}, \bibinfo{person}{David Schnurr}, \bibinfo{person}{John Schulman}, \bibinfo{person}{Daniel Selsam},
  \bibinfo{person}{Kyla Sheppard}, \bibinfo{person}{Toki Sherbakov}, \bibinfo{person}{Jessica Shieh}, \bibinfo{person}{Sarah Shoker}, \bibinfo{person}{Pranav Shyam}, \bibinfo{person}{Szymon Sidor}, \bibinfo{person}{Eric Sigler}, \bibinfo{person}{Maddie Simens}, \bibinfo{person}{Jordan Sitkin}, \bibinfo{person}{Katarina Slama}, \bibinfo{person}{Ian Sohl}, \bibinfo{person}{Benjamin Sokolowsky}, \bibinfo{person}{Yang Song}, \bibinfo{person}{Natalie Staudacher}, \bibinfo{person}{Felipe~Petroski Such}, \bibinfo{person}{Natalie Summers}, \bibinfo{person}{Ilya Sutskever}, \bibinfo{person}{Jie Tang}, \bibinfo{person}{Nikolas Tezak}, \bibinfo{person}{Madeleine~B. Thompson}, \bibinfo{person}{Phil Tillet}, \bibinfo{person}{Amin Tootoonchian}, \bibinfo{person}{Elizabeth Tseng}, \bibinfo{person}{Preston Tuggle}, \bibinfo{person}{Nick Turley}, \bibinfo{person}{Jerry Tworek}, \bibinfo{person}{Juan Felipe~Cerón Uribe}, \bibinfo{person}{Andrea Vallone}, \bibinfo{person}{Arun Vijayvergiya}, \bibinfo{person}{Chelsea Voss},
  \bibinfo{person}{Carroll Wainwright}, \bibinfo{person}{Justin~Jay Wang}, \bibinfo{person}{Alvin Wang}, \bibinfo{person}{Ben Wang}, \bibinfo{person}{Jonathan Ward}, \bibinfo{person}{Jason Wei}, \bibinfo{person}{CJ Weinmann}, \bibinfo{person}{Akila Welihinda}, \bibinfo{person}{Peter Welinder}, \bibinfo{person}{Jiayi Weng}, \bibinfo{person}{Lilian Weng}, \bibinfo{person}{Matt Wiethoff}, \bibinfo{person}{Dave Willner}, \bibinfo{person}{Clemens Winter}, \bibinfo{person}{Samuel Wolrich}, \bibinfo{person}{Hannah Wong}, \bibinfo{person}{Lauren Workman}, \bibinfo{person}{Sherwin Wu}, \bibinfo{person}{Jeff Wu}, \bibinfo{person}{Michael Wu}, \bibinfo{person}{Kai Xiao}, \bibinfo{person}{Tao Xu}, \bibinfo{person}{Sarah Yoo}, \bibinfo{person}{Kevin Yu}, \bibinfo{person}{Qiming Yuan}, \bibinfo{person}{Wojciech Zaremba}, \bibinfo{person}{Rowan Zellers}, \bibinfo{person}{Chong Zhang}, \bibinfo{person}{Marvin Zhang}, \bibinfo{person}{Shengjia Zhao}, \bibinfo{person}{Tianhao Zheng}, \bibinfo{person}{Juntang Zhuang},
  \bibinfo{person}{William Zhuk}, {and} \bibinfo{person}{Barret Zoph}.} \bibinfo{year}{2024}\natexlab{}.
\newblock \bibinfo{title}{GPT-4 Technical Report}.
\newblock
\newblock
\showeprint[arxiv]{2303.08774}~[cs.CL]
\urldef\tempurl%
\url{https://arxiv.org/abs/2303.08774}
\showURL{%
\tempurl}


\bibitem[Raffel et~al\mbox{.}(2023)]%
        {raffel2023exploringlimitstransferlearning}
\bibfield{author}{\bibinfo{person}{Colin Raffel}, \bibinfo{person}{Noam Shazeer}, \bibinfo{person}{Adam Roberts}, \bibinfo{person}{Katherine Lee}, \bibinfo{person}{Sharan Narang}, \bibinfo{person}{Michael Matena}, \bibinfo{person}{Yanqi Zhou}, \bibinfo{person}{Wei Li}, {and} \bibinfo{person}{Peter~J. Liu}.} \bibinfo{year}{2023}\natexlab{}.
\newblock \bibinfo{title}{Exploring the Limits of Transfer Learning with a Unified Text-to-Text Transformer}.
\newblock
\newblock
\showeprint[arxiv]{1910.10683}~[cs.LG]
\urldef\tempurl%
\url{https://arxiv.org/abs/1910.10683}
\showURL{%
\tempurl}


\bibitem[Ren et~al\mbox{.}(2024)]%
        {ren2024robustsemanticsbasedwatermarklarge}
\bibfield{author}{\bibinfo{person}{Jie Ren}, \bibinfo{person}{Han Xu}, \bibinfo{person}{Yiding Liu}, \bibinfo{person}{Yingqian Cui}, \bibinfo{person}{Shuaiqiang Wang}, \bibinfo{person}{Dawei Yin}, {and} \bibinfo{person}{Jiliang Tang}.} \bibinfo{year}{2024}\natexlab{}.
\newblock \bibinfo{title}{A Robust Semantics-based Watermark for Large Language Model against Paraphrasing}.
\newblock
\newblock
\showeprint[arxiv]{2311.08721}~[cs.CR]
\urldef\tempurl%
\url{https://arxiv.org/abs/2311.08721}
\showURL{%
\tempurl}


\bibitem[Santurkar et~al\mbox{.}(2021)]%
        {santurkar2021editingclassifierrewritingprediction}
\bibfield{author}{\bibinfo{person}{Shibani Santurkar}, \bibinfo{person}{Dimitris Tsipras}, \bibinfo{person}{Mahalaxmi Elango}, \bibinfo{person}{David Bau}, \bibinfo{person}{Antonio Torralba}, {and} \bibinfo{person}{Aleksander Madry}.} \bibinfo{year}{2021}\natexlab{}.
\newblock \bibinfo{title}{Editing a classifier by rewriting its prediction rules}.
\newblock
\newblock
\showeprint[arxiv]{2112.01008}~[cs.LG]
\urldef\tempurl%
\url{https://arxiv.org/abs/2112.01008}
\showURL{%
\tempurl}


\bibitem[Shirali-Shahreza and Shirali-Shahreza(2008)]%
        {shirali2008synonymsteganography}
\bibfield{author}{\bibinfo{person}{M.~Hassan Shirali-Shahreza} {and} \bibinfo{person}{Mohammad Shirali-Shahreza}.} \bibinfo{year}{2008}\natexlab{}.
\newblock \showarticletitle{A New Synonym Text Steganography}. In \bibinfo{booktitle}{\emph{2008 International Conference on Intelligent Information Hiding and Multimedia Signal Processing}}. \bibinfo{pages}{1524--1526}.
\newblock
\urldef\tempurl%
\url{https://doi.org/10.1109/IIH-MSP.2008.6}
\showDOI{\tempurl}


\bibitem[Singh and Zou(2023)]%
        {singh2023newevaluationmetricscapture}
\bibfield{author}{\bibinfo{person}{Karanpartap Singh} {and} \bibinfo{person}{James Zou}.} \bibinfo{year}{2023}\natexlab{}.
\newblock \bibinfo{title}{New Evaluation Metrics Capture Quality Degradation due to LLM Watermarking}.
\newblock
\newblock
\showeprint[arxiv]{2312.02382}~[cs.CL]
\urldef\tempurl%
\url{https://arxiv.org/abs/2312.02382}
\showURL{%
\tempurl}


\bibitem[Sung et~al\mbox{.}(2021)]%
        {sung2021trainingneuralnetworksfixed}
\bibfield{author}{\bibinfo{person}{Yi-Lin Sung}, \bibinfo{person}{Varun Nair}, {and} \bibinfo{person}{Colin Raffel}.} \bibinfo{year}{2021}\natexlab{}.
\newblock \bibinfo{title}{Training Neural Networks with Fixed Sparse Masks}.
\newblock
\newblock
\showeprint[arxiv]{2111.09839}~[cs.LG]
\urldef\tempurl%
\url{https://arxiv.org/abs/2111.09839}
\showURL{%
\tempurl}


\bibitem[Takezawa et~al\mbox{.}(2023)]%
        {takezawa2023necessarysufficientwatermarklarge}
\bibfield{author}{\bibinfo{person}{Yuki Takezawa}, \bibinfo{person}{Ryoma Sato}, \bibinfo{person}{Han Bao}, \bibinfo{person}{Kenta Niwa}, {and} \bibinfo{person}{Makoto Yamada}.} \bibinfo{year}{2023}\natexlab{}.
\newblock \bibinfo{title}{Necessary and Sufficient Watermark for Large Language Models}.
\newblock
\newblock
\showeprint[arxiv]{2310.00833}~[cs.CL]
\urldef\tempurl%
\url{https://arxiv.org/abs/2310.00833}
\showURL{%
\tempurl}


\bibitem[Touvron et~al\mbox{.}(2023)]%
        {touvron2023llama2openfoundation}
\bibfield{author}{\bibinfo{person}{Hugo Touvron}, \bibinfo{person}{Louis Martin}, \bibinfo{person}{Kevin Stone}, \bibinfo{person}{Peter Albert}, \bibinfo{person}{Amjad Almahairi}, \bibinfo{person}{Yasmine Babaei}, \bibinfo{person}{Nikolay Bashlykov}, \bibinfo{person}{Soumya Batra}, \bibinfo{person}{Prajjwal Bhargava}, \bibinfo{person}{Shruti Bhosale}, \bibinfo{person}{Dan Bikel}, \bibinfo{person}{Lukas Blecher}, \bibinfo{person}{Cristian~Canton Ferrer}, \bibinfo{person}{Moya Chen}, \bibinfo{person}{Guillem Cucurull}, \bibinfo{person}{David Esiobu}, \bibinfo{person}{Jude Fernandes}, \bibinfo{person}{Jeremy Fu}, \bibinfo{person}{Wenyin Fu}, \bibinfo{person}{Brian Fuller}, \bibinfo{person}{Cynthia Gao}, \bibinfo{person}{Vedanuj Goswami}, \bibinfo{person}{Naman Goyal}, \bibinfo{person}{Anthony Hartshorn}, \bibinfo{person}{Saghar Hosseini}, \bibinfo{person}{Rui Hou}, \bibinfo{person}{Hakan Inan}, \bibinfo{person}{Marcin Kardas}, \bibinfo{person}{Viktor Kerkez}, \bibinfo{person}{Madian Khabsa},
  \bibinfo{person}{Isabel Kloumann}, \bibinfo{person}{Artem Korenev}, \bibinfo{person}{Punit~Singh Koura}, \bibinfo{person}{Marie-Anne Lachaux}, \bibinfo{person}{Thibaut Lavril}, \bibinfo{person}{Jenya Lee}, \bibinfo{person}{Diana Liskovich}, \bibinfo{person}{Yinghai Lu}, \bibinfo{person}{Yuning Mao}, \bibinfo{person}{Xavier Martinet}, \bibinfo{person}{Todor Mihaylov}, \bibinfo{person}{Pushkar Mishra}, \bibinfo{person}{Igor Molybog}, \bibinfo{person}{Yixin Nie}, \bibinfo{person}{Andrew Poulton}, \bibinfo{person}{Jeremy Reizenstein}, \bibinfo{person}{Rashi Rungta}, \bibinfo{person}{Kalyan Saladi}, \bibinfo{person}{Alan Schelten}, \bibinfo{person}{Ruan Silva}, \bibinfo{person}{Eric~Michael Smith}, \bibinfo{person}{Ranjan Subramanian}, \bibinfo{person}{Xiaoqing~Ellen Tan}, \bibinfo{person}{Binh Tang}, \bibinfo{person}{Ross Taylor}, \bibinfo{person}{Adina Williams}, \bibinfo{person}{Jian~Xiang Kuan}, \bibinfo{person}{Puxin Xu}, \bibinfo{person}{Zheng Yan}, \bibinfo{person}{Iliyan Zarov}, \bibinfo{person}{Yuchen
  Zhang}, \bibinfo{person}{Angela Fan}, \bibinfo{person}{Melanie Kambadur}, \bibinfo{person}{Sharan Narang}, \bibinfo{person}{Aurelien Rodriguez}, \bibinfo{person}{Robert Stojnic}, \bibinfo{person}{Sergey Edunov}, {and} \bibinfo{person}{Thomas Scialom}.} \bibinfo{year}{2023}\natexlab{}.
\newblock \bibinfo{title}{Llama 2: Open Foundation and Fine-Tuned Chat Models}.
\newblock
\newblock
\showeprint[arxiv]{2307.09288}~[cs.CL]
\urldef\tempurl%
\url{https://arxiv.org/abs/2307.09288}
\showURL{%
\tempurl}


\bibitem[Tu et~al\mbox{.}(2024)]%
        {tu2024waterbenchholisticevaluationwatermarks}
\bibfield{author}{\bibinfo{person}{Shangqing Tu}, \bibinfo{person}{Yuliang Sun}, \bibinfo{person}{Yushi Bai}, \bibinfo{person}{Jifan Yu}, \bibinfo{person}{Lei Hou}, {and} \bibinfo{person}{Juanzi Li}.} \bibinfo{year}{2024}\natexlab{}.
\newblock \bibinfo{title}{WaterBench: Towards Holistic Evaluation of Watermarks for Large Language Models}.
\newblock
\newblock
\showeprint[arxiv]{2311.07138}~[cs.CL]
\urldef\tempurl%
\url{https://arxiv.org/abs/2311.07138}
\showURL{%
\tempurl}


\bibitem[Ueoka et~al\mbox{.}(2021)]%
        {ueoka2021frustratinglyeasyeditbasedlinguistic}
\bibfield{author}{\bibinfo{person}{Honai Ueoka}, \bibinfo{person}{Yugo Murawaki}, {and} \bibinfo{person}{Sadao Kurohashi}.} \bibinfo{year}{2021}\natexlab{}.
\newblock \bibinfo{title}{Frustratingly Easy Edit-based Linguistic Steganography with a Masked Language Model}.
\newblock
\newblock
\showeprint[arxiv]{2104.09833}~[cs.CL]
\urldef\tempurl%
\url{https://arxiv.org/abs/2104.09833}
\showURL{%
\tempurl}


\bibitem[Vaswani et~al\mbox{.}(2023)]%
        {vaswani2023attentionneed}
\bibfield{author}{\bibinfo{person}{Ashish Vaswani}, \bibinfo{person}{Noam Shazeer}, \bibinfo{person}{Niki Parmar}, \bibinfo{person}{Jakob Uszkoreit}, \bibinfo{person}{Llion Jones}, \bibinfo{person}{Aidan~N. Gomez}, \bibinfo{person}{Lukasz Kaiser}, {and} \bibinfo{person}{Illia Polosukhin}.} \bibinfo{year}{2023}\natexlab{}.
\newblock \bibinfo{title}{Attention Is All You Need}.
\newblock
\newblock
\showeprint[arxiv]{1706.03762}~[cs.CL]
\urldef\tempurl%
\url{https://arxiv.org/abs/1706.03762}
\showURL{%
\tempurl}


\bibitem[Venugopal et~al\mbox{.}(2011)]%
        {venugopal-etal-2011-watermarking}
\bibfield{author}{\bibinfo{person}{Ashish Venugopal}, \bibinfo{person}{Jakob Uszkoreit}, \bibinfo{person}{David Talbot}, \bibinfo{person}{Franz Och}, {and} \bibinfo{person}{Juri Ganitkevitch}.} \bibinfo{year}{2011}\natexlab{}.
\newblock \showarticletitle{Watermarking the Outputs of Structured Prediction with an application in Statistical Machine Translation.}. In \bibinfo{booktitle}{\emph{Proceedings of the 2011 Conference on Empirical Methods in Natural Language Processing}}, \bibfield{editor}{\bibinfo{person}{Regina Barzilay} {and} \bibinfo{person}{Mark Johnson}} (Eds.). \bibinfo{publisher}{Association for Computational Linguistics}, \bibinfo{address}{Edinburgh, Scotland, UK.}, \bibinfo{pages}{1363--1372}.
\newblock
\urldef\tempurl%
\url{https://aclanthology.org/D11-1126}
\showURL{%
\tempurl}


\bibitem[Wang et~al\mbox{.}(2024)]%
        {wang2024codablewatermarkinginjectingmultibits}
\bibfield{author}{\bibinfo{person}{Lean Wang}, \bibinfo{person}{Wenkai Yang}, \bibinfo{person}{Deli Chen}, \bibinfo{person}{Hao Zhou}, \bibinfo{person}{Yankai Lin}, \bibinfo{person}{Fandong Meng}, \bibinfo{person}{Jie Zhou}, {and} \bibinfo{person}{Xu Sun}.} \bibinfo{year}{2024}\natexlab{}.
\newblock \bibinfo{title}{Towards Codable Watermarking for Injecting Multi-bits Information to LLMs}.
\newblock
\newblock
\showeprint[arxiv]{2307.15992}~[cs.CL]
\urldef\tempurl%
\url{https://arxiv.org/abs/2307.15992}
\showURL{%
\tempurl}


\bibitem[Wang et~al\mbox{.}(2023)]%
        {wang2023comprehensivesurveyforgettingdeep}
\bibfield{author}{\bibinfo{person}{Zhenyi Wang}, \bibinfo{person}{Enneng Yang}, \bibinfo{person}{Li Shen}, {and} \bibinfo{person}{Heng Huang}.} \bibinfo{year}{2023}\natexlab{}.
\newblock \bibinfo{title}{A Comprehensive Survey of Forgetting in Deep Learning Beyond Continual Learning}.
\newblock
\newblock
\showeprint[arxiv]{2307.09218}~[cs.LG]
\urldef\tempurl%
\url{https://arxiv.org/abs/2307.09218}
\showURL{%
\tempurl}


\bibitem[Welleck et~al\mbox{.}(2019)]%
        {welleck2019neuraltextgenerationunlikelihood}
\bibfield{author}{\bibinfo{person}{Sean Welleck}, \bibinfo{person}{Ilia Kulikov}, \bibinfo{person}{Stephen Roller}, \bibinfo{person}{Emily Dinan}, \bibinfo{person}{Kyunghyun Cho}, {and} \bibinfo{person}{Jason Weston}.} \bibinfo{year}{2019}\natexlab{}.
\newblock \bibinfo{title}{Neural Text Generation with Unlikelihood Training}.
\newblock
\newblock
\showeprint[arxiv]{1908.04319}~[cs.LG]
\urldef\tempurl%
\url{https://arxiv.org/abs/1908.04319}
\showURL{%
\tempurl}


\bibitem[Wilson et~al\mbox{.}(2014)]%
        {DBLP:conf/mediaforensics/WilsonBK14}
\bibfield{author}{\bibinfo{person}{Alex Wilson}, \bibinfo{person}{Phil Blunsom}, {and} \bibinfo{person}{Andrew~D. Ker}.} \bibinfo{year}{2014}\natexlab{}.
\newblock \showarticletitle{Linguistic steganography on Twitter: hierarchical language modeling with manual interaction}. In \bibinfo{booktitle}{\emph{Media Watermarking, Security, and Forensics 2014, San Francisco, CA, USA, February 2, 2014, Proceedings}} \emph{(\bibinfo{series}{{SPIE} Proceedings}, Vol.~\bibinfo{volume}{9028})}, \bibfield{editor}{\bibinfo{person}{Adnan~M. Alattar}, \bibinfo{person}{Nasir~D. Memon}, {and} \bibinfo{person}{Chad Heitzenrater}} (Eds.). \bibinfo{publisher}{{SPIE}}, \bibinfo{pages}{902803}.
\newblock
\urldef\tempurl%
\url{https://doi.org/10.1117/12.2039213}
\showDOI{\tempurl}


\bibitem[Wilson and Ker(2016)]%
        {wilson2016steganograph}
\bibfield{author}{\bibinfo{person}{Alex Wilson} {and} \bibinfo{person}{AndrewD Ker}.} \bibinfo{year}{2016}\natexlab{}.
\newblock \showarticletitle{Avoiding detection on twitter: embedding strategies for linguistic steganography}.
\newblock \bibinfo{journal}{\emph{Electronic Imaging}}  \bibinfo{volume}{2016} (\bibinfo{date}{02} \bibinfo{year}{2016}), \bibinfo{pages}{1--9}.
\newblock
\urldef\tempurl%
\url{https://doi.org/10.2352/ISSN.2470-1173.2016.8.MWSF-074}
\showDOI{\tempurl}


\bibitem[Wu et~al\mbox{.}(2024)]%
        {wu2024resilientaccessibledistributionpreservingwatermark}
\bibfield{author}{\bibinfo{person}{Yihan Wu}, \bibinfo{person}{Zhengmian Hu}, \bibinfo{person}{Junfeng Guo}, \bibinfo{person}{Hongyang Zhang}, {and} \bibinfo{person}{Heng Huang}.} \bibinfo{year}{2024}\natexlab{}.
\newblock \bibinfo{title}{A Resilient and Accessible Distribution-Preserving Watermark for Large Language Models}.
\newblock
\newblock
\showeprint[arxiv]{2310.07710}~[cs.CR]
\urldef\tempurl%
\url{https://arxiv.org/abs/2310.07710}
\showURL{%
\tempurl}


\bibitem[Yang et~al\mbox{.}(2021)]%
        {yang2021tracingtextprovenancecontextaware}
\bibfield{author}{\bibinfo{person}{Xi Yang}, \bibinfo{person}{Jie Zhang}, \bibinfo{person}{Kejiang Chen}, \bibinfo{person}{Weiming Zhang}, \bibinfo{person}{Zehua Ma}, \bibinfo{person}{Feng Wang}, {and} \bibinfo{person}{Nenghai Yu}.} \bibinfo{year}{2021}\natexlab{}.
\newblock \bibinfo{title}{Tracing Text Provenance via Context-Aware Lexical Substitution}.
\newblock
\newblock
\showeprint[arxiv]{2112.07873}~[cs.CR]
\urldef\tempurl%
\url{https://arxiv.org/abs/2112.07873}
\showURL{%
\tempurl}


\bibitem[Yang et~al\mbox{.}(2024)]%
        {yang2024revisitextendenhancehessianfree}
\bibfield{author}{\bibinfo{person}{Ziao Yang}, \bibinfo{person}{Han Yue}, \bibinfo{person}{Jian Chen}, {and} \bibinfo{person}{Hongfu Liu}.} \bibinfo{year}{2024}\natexlab{}.
\newblock \bibinfo{title}{Revisit, Extend, and Enhance Hessian-Free Influence Functions}.
\newblock
\newblock
\showeprint[arxiv]{2405.17490}~[cs.LG]
\urldef\tempurl%
\url{https://arxiv.org/abs/2405.17490}
\showURL{%
\tempurl}


\bibitem[Zhao et~al\mbox{.}(2023)]%
        {zhao2023provablerobustwatermarkingaigenerated}
\bibfield{author}{\bibinfo{person}{Xuandong Zhao}, \bibinfo{person}{Prabhanjan Ananth}, \bibinfo{person}{Lei Li}, {and} \bibinfo{person}{Yu-Xiang Wang}.} \bibinfo{year}{2023}\natexlab{}.
\newblock \bibinfo{title}{Provable Robust Watermarking for AI-Generated Text}.
\newblock
\newblock
\showeprint[arxiv]{2306.17439}~[cs.CL]
\urldef\tempurl%
\url{https://arxiv.org/abs/2306.17439}
\showURL{%
\tempurl}


\bibitem[Zhao et~al\mbox{.}(2024)]%
        {zhao2024permuteandflipoptimallyrobustwatermarkable}
\bibfield{author}{\bibinfo{person}{Xuandong Zhao}, \bibinfo{person}{Lei Li}, {and} \bibinfo{person}{Yu-Xiang Wang}.} \bibinfo{year}{2024}\natexlab{}.
\newblock \bibinfo{title}{Permute-and-Flip: An optimally robust and watermarkable decoder for LLMs}.
\newblock
\newblock
\showeprint[arxiv]{2402.05864}~[cs.CL]
\urldef\tempurl%
\url{https://arxiv.org/abs/2402.05864}
\showURL{%
\tempurl}


\bibitem[Zhu et~al\mbox{.}(2024)]%
        {zhu2024duwakdualwatermarkslarge}
\bibfield{author}{\bibinfo{person}{Chaoyi Zhu}, \bibinfo{person}{Jeroen Galjaard}, \bibinfo{person}{Pin-Yu Chen}, {and} \bibinfo{person}{Lydia~Y. Chen}.} \bibinfo{year}{2024}\natexlab{}.
\newblock \bibinfo{title}{Duwak: Dual Watermarks in Large Language Models}.
\newblock
\newblock
\showeprint[arxiv]{2403.13000}~[cs.LG]
\urldef\tempurl%
\url{https://arxiv.org/abs/2403.13000}
\showURL{%
\tempurl}


\bibitem[Ziegler et~al\mbox{.}(2019)]%
        {ziegler2019neurallinguisticsteganography}
\bibfield{author}{\bibinfo{person}{Zachary~M. Ziegler}, \bibinfo{person}{Yuntian Deng}, {and} \bibinfo{person}{Alexander~M. Rush}.} \bibinfo{year}{2019}\natexlab{}.
\newblock \bibinfo{title}{Neural Linguistic Steganography}.
\newblock
\newblock
\showeprint[arxiv]{1909.01496}~[cs.CL]
\urldef\tempurl%
\url{https://arxiv.org/abs/1909.01496}
\showURL{%
\tempurl}


\end{thebibliography}

\appendix
\newpage
\section{Fine-tuned models' watermarking distillation setup}
\label{appendix:motivation_finetuning_setup}
We use Neuralmagic Llama-2-7B-gsm8k~\citep{agarwalla2024enablinghighsparsityfoundationalllama} as both the teacher and student models. Mathematics is selected as the fine-tuned capability, with GSM8K~\citep{cobbe2021gsm8kmath} serving as the fine-tuning dataset.

Due to the low-entropy nature of mathematics questions, which can interfere with the watermarking process, we use Chain-of-Thought (CoT) prompting on the fine-tuning samples to expand the entropy space and improve detectability, ensuring that the model demonstrates both fine-tuned capability and watermarking simultaneously.

For watermarking, we select the schemes kgw-k0-gamma-0.25-delta-2 and aar-k3, as they are relatively easier for the model to learn from. 
We will now introduce the specific setups for the three different approaches described in $\S$ 2.1.

\paragraph{Distilling a fine-tuned model with watermarked content}
We use a math-fine-tuned model as the student model and Llama 2 7B~\citep{touvron2023llama2openfoundation} as the teacher model. 
The distillation process utilizes the OpenWebText dataset~\citep{Gokaslan2019OpenWeb} for 1,000 steps.

The batch size is set to 64, the sequence length to 512, and the maximum learning rate to $1 \times 10^{-5}$, with a cosine learning rate decay and a linear warmup during the first 200 steps. Each training session takes approximately 2 hours on 4 NVIDIA A100 80GB GPUs.

\paragraph{Fine-tuning a distilled model that already contains a watermark}
We used a watermarked fine-tuned Llama 2 7B model and fine-tuned it further on GSM8K to enhance its math capabilities. 
The total training consisted of 129 steps with a batch size of 64 sequences and a sequence length of 256 tokens.

The maximum learning rate was set to $1 \times 10^{-5}$, with a cosine learning rate decay and a linear warmup over the first 20 steps. 
We employed the AdamW optimizer with $(\beta_1, \beta_2) = (0.9, 0.999)$ and no weight decay. 
Each training run took approximately 50 minutes on 4 NVIDIA A100 80GB GPUs.

\paragraph{Fine-tuning a base model using a watermarked fine-tuning dataset.}
First, we generated watermarked samples of 256 tokens using a 50-token prefix from GSM8K as the prompt. 
These watermarked generations were filtered based on the correctness of their final answers, resulting in 2,632 correct samples, which were used as training data for distilling the Llama-2-7B-gsm8k model.

Next, we fine-tuned Llama-2-7B-gsm8k on the watermarked samples for 3 epochs, with 43 steps per epoch, using a batch size of 64 sequences and a sequence length of 256 tokens. 
The maximum learning rate was set to $1 \times 10^{-5}$, with a cosine learning rate decay and a linear warmup over the first 20 steps. 
We used the AdamW optimizer with $(\beta_1, \beta_2) = (0.9, 0.999)$ and no weight decay. 
Each training run took approximately 60 minutes on 4 NVIDIA A100 80GB GPUs.

In our experiments, we use Math, QA, and instruction-tuning models based on Llama-2-7B. 
Since~\cite{gu2024learnabilitywatermarkslanguagemodels} provides pre-watermarked distilled models for Llama-2-7B, the watermarking process for these models incurs no additional training cost. 
Even if pre-watermarked models are unavailable, WAPITI requires only a single watermark distillation on the base model to watermark fine-tuned models of the same type. 
In contrast, vanilla watermark distillation necessitates a separate distillation process for each fine-tuned model, highlighting the efficiency of WAPITI.


\section{Detail definition for watermark schemes}
\label{appendix:detail_watermark_definition}
In this section, we will provide rigid definitions of watermark schemes used in this work: KGW~\citep{kirchenbauer2024watermarklargelanguagemodels} and AAR~\citep{aaronson2023watermarking}.

\paragraph{KGW} 
For the KGW watermark, we use the same notation as described in the main text: \( \mathcal{W}^{KGW} \) represents the watermarking algorithm, \( f_{\boldsymbol{\theta}}(\cdot \mid \boldsymbol{x}) \) denotes the next-token probability, and \( \phi \) is the watermark key. 
The hyperparameters \( k, \gamma, \delta \) are specific to KGW, where \( k \) defines how many preceding tokens are used to compute the corresponding green list of the next token, \( \gamma \) indicates the proportion of the vocabulary in the green list, and \( \delta \) refers to the watermark shift applied to the tokens in the green list.
The full logit generation process for KGW is defined as:
\begin{align}
    f_{\boldsymbol\theta}^{KGW}(\boldsymbol x, \phi, k, \gamma, \delta) 
    &= \text{softmax}\left(\log(f_{\boldsymbol \theta}(\cdot\, | x)) + \right. \nonumber \\
    & \quad \left. \delta \cdot \mathcal{W}^{KGW}(x_{i-k}, \cdots, x_{i-1};\phi;\gamma;|\mathcal{V}|)\right)
\end{align}
Here $\mathcal{W}^{KGW}$ is a hash function that generates the green token list mask according to the watermark hyperparameter. 

The detection of the KGW watermark is:
\begin{align}
    \mathcal{D}(\boldsymbol{x},\phi,\gamma) = 1 - {Bino}\underbrace{\left(\sum_{t=0}^{len(x} x_t\cdot \mathcal{W}^{KGW}(x_{t-k}, \cdots, x_{t-1};\phi;\gamma;|\mathcal{V}|) \right)}_{\text{number of green list tokens in $\boldsymbol x$}}
\end{align}
Where the term within the parenthesis is calculating how many tokens with the green list and $Bino$ here refers to the cumulative distribution function for binomial distributed random variables.

\paragraph{AAR}
For the AAR watermark, we use the same notation as well.
\(\mathcal{W}^{AAR} \) represents the watermarking algorithm, \( f_{\boldsymbol{\theta}}(\cdot \mid \boldsymbol{x}) \) denotes the next-token probability, and \( \phi \) is the watermark key.
AAR only has one hyperparameter $k$ that denotes how many preceding tokens are used to compute the score sequence $\boldsymbol r_i$.
\begin{align}
   \boldsymbol r_i = \mathcal{W}^{AAR}(x_{i-k}, \cdots, x_{i-1}, \phi) \sim \text{Unif}(0,1)^{|\mathcal{V}|}
\end{align}
The full token sampling process for AAR is defined as:
\begin{align}
   x_i^{AAR}= (\argmax_{j\in|\mathcal{V}|}(\log(f_{\boldsymbol \theta}(\cdot\, | x))^j - \log(-\log(r_i^j))   
\end{align}
The detection of the AAR watermark is:
\begin{align}
    \mathcal{D}^{AAR}(\boldsymbol{x}, \phi, \gamma) &= 1 - \Gamma(len(\boldsymbol{x}) - k, 1) \nonumber \cdot \\
    &\quad \left( \sum_{t=0}^{len(\boldsymbol{x})} - \log\left( 1 - \underbrace{\mathcal{W}^{AAR}(x_{i-k}, \dots, x_{i-1}, \phi)_{x_t}}_{\text{corresponding score of } x_i} \right) \right)
\end{align}

\section{Preliminary for n-gram distribution analysis}
\label{appendix:ngram_analysis}
\cite{gu2024learnabilitywatermarkslanguagemodels} has demonstrated that the distilled model achieves satisfactory watermarking performance.
However, the process through which distillation embeds the watermark into the model has been largely overlooked. 
Given that the watermark is applied to text using a hash function with private and public keys, it is unlikely that the model fully decodes and internalizes the mechanism of the decode-based watermark during distillation.
\begin{figure}{r}
    \centering
    \includegraphics[width=0.7\linewidth]{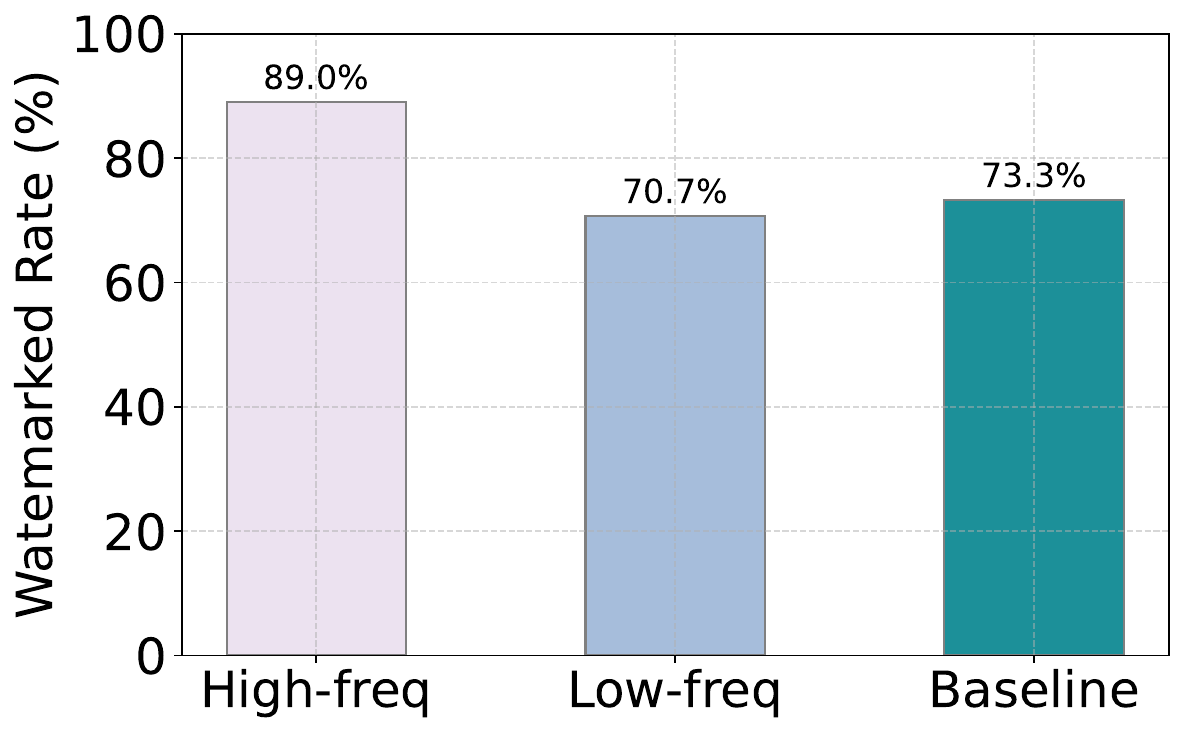}
    \caption{The results show the proportion of watermarked content generated from prefixes of high- and low-frequency watermarked $n$-grams in the distillation data. The baseline uses prefixes from unwatermarked $n$-grams in the same data.}
    \label{fig:ngram_distillation_watermarked_rate}
    \vspace{-10pt}
\end{figure}
We hypothesize that the core knowledge acquired during distillation relates to n-grams. To test this, we conduct experiments using KGW and AAR as decoding-based watermarks, with the Llama model family for consistency.

First, the use of n-grams is grounded in theoretical reasoning. As defined for KGW and AAR in~\ref{appendix:detail_watermark_definition}, detecting \(x_i\) relies only on \(x_{i-k}, \dots, x_{i-1}\), allowing the partitioning of a sentence into \((k+1)\)-grams for detection.

Next, we investigate how watermark distillation affects the n-gram distribution in model outputs. Using 256,000 samples of length 50 from the training data of the k1-gamma0.25-delta2 watermark, we tokenize the data into bigrams for analysis (since \(k=1\)). We then compare high- and low-frequency watermarked bigrams to see if the model generates corresponding content. Unwatermarked bigrams serve as the baseline. Results in Figure~\ref{fig:ngram_distillation_watermarked_rate} show that the model consistently generates watermarked content for high-frequency bigrams, while for low-frequency bigrams, the behavior resembles the baseline with fewer watermarked outputs.

This result validates that the model learns the watermarking strategy at the n-gram level, confirming that analyzing the model from an n-gram perspective is appropriate.

\section{Fine-tuned model choices in the main experiment}
\label{appendix:finetuned_model_choice}
For Llama models, we choose alpaca-7b-reproduced-llama-2~\citep{safe-rlhf} as QA fine-tuned model, Llama-2-7b-gsm8k~\citep{agarwalla2024enablinghighsparsityfoundationalllama} as math fine-tuned model and Llama-2-7b-chat-hf~\cite{touvron2023llama2openfoundation} as instruction fine-tuned model. 
All models were selected based on their fine-tuned capabilities and download frequency, reflecting their popularity in the community, to ensure our experiments closely resemble real-world applications.
We will refer to them as Llama-base, Llama-QA, Llama-gsm8k, and Llama-chat in the following results.
For Pythia models, because of the ability limit of Pythia-1.4B, we only choose Pythia-1.4B-sft~\citep{lambdalabs_pythia_1.4b_2024}, which will be referred to as Pythia-base and Pythia-chat.

\section{\Ours{}'s robustness to post-processing}
\label{appendix:robustness_analysis}
\paragraph{Text Edit}
The text editing experiment focuses on whether watermarked text remains detectable after randomly corrupting a certain portion of tokens. 
We test all hyperparameter sets of KGW and AAR watermarking. 
Specifically,  \( k = \{0, 1, 2\} \), \( \gamma = 0.25 \), and \( \delta = \{1, 2\} \) for KGW; and for AAR, we use \( k = \{2, 3, 4\} \).
The generations are taken from the \Ours{} watermarked model described in $\S$4.2.

\begin{figure}[t]
    \centering
    \begin{minipage}{0.35\textwidth} 
        \centering
        \includegraphics[width=\textwidth]{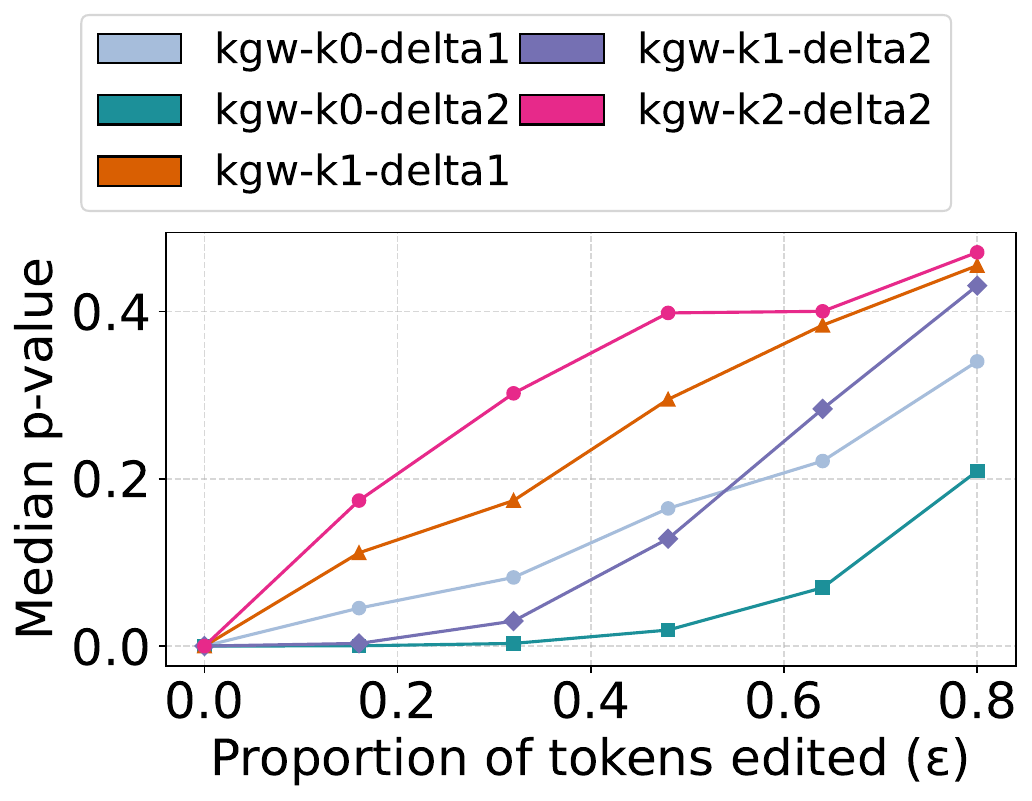} 
    \end{minipage}
    \hfill 
    \begin{minipage}{0.35\textwidth}
        \centering
        \includegraphics[width=\textwidth]{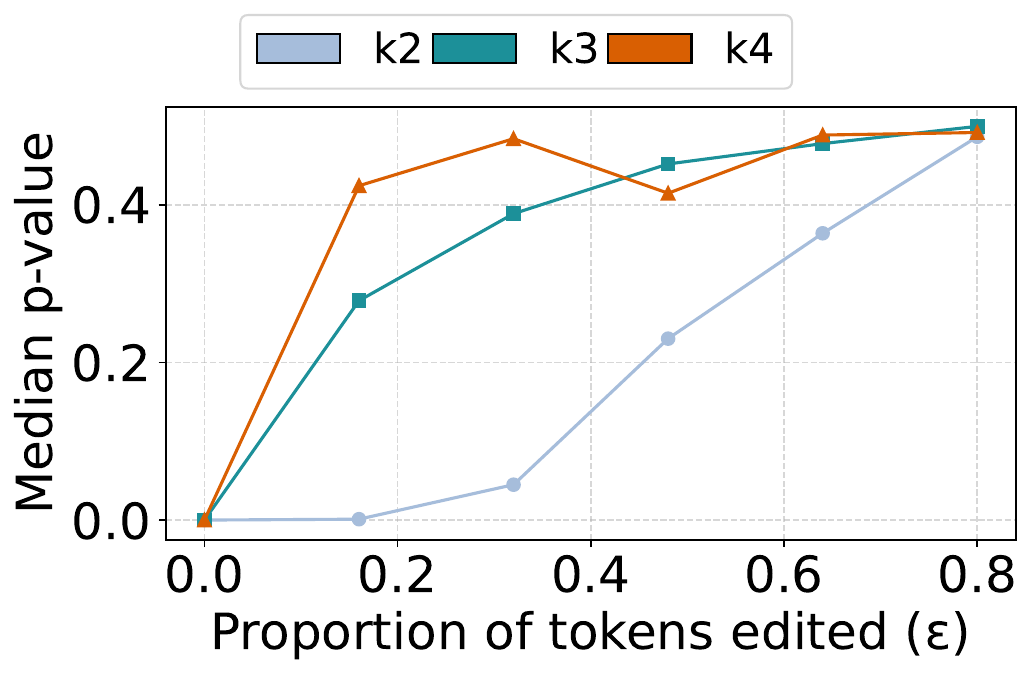} 
    \end{minipage}
    \caption{Watermark detection p-values for generations from the KGW watermark after text edits at various proportions $\epsilon$. 
    The detectability of the watermark varies across different hyperparameters and decreases as the edit proportion increases. 
    Overall, \Ours{} watermarks exhibit robustness to mild text corruption.}
    \label{fig:text_edit_result}
\end{figure}

First, we set the edit proportion $\epsilon = \{0, 0.16, 0.32, 0.48, 0.64, 0.8\}$. Then, for each generated sample, we randomly select a proportion $\epsilon$ of tokens and replace each with a random token drawn uniformly from the tokenizer's vocabulary. Finally, we compute the median p-value of the edited sequences to assess their detectability.

As shown in Figure~\ref{fig:text_edit_result}, detectability of text remains robust to text edits when $\epsilon$ up to 20\%.
A higher corruption rate would lead to substantial decay of watermark detectability.
Interestingly, the robustness of text editing appears to be closely related to the window size of the watermarking method. Specifically, smaller window sizes ($k$ in both KGW and AAR) demonstrate greater robustness to token corruption. 
This observation aligns with the results presented in Section~\ref{appendix:coeff_analysis}.

\end{document}